\documentclass[12pt]{article}
\usepackage{amsmath}
\usepackage{amsfonts}
\usepackage{amssymb}
\usepackage{mathptmx}
\usepackage{slashed}
\usepackage{latexsym}
\usepackage{graphicx}
\usepackage{color}
\newcommand{\newc}{\newcommand}
\newc{\be}{\begin{equation}}
\newc{\ee}{\end{equation}}
\newc{\bea}{\begin{eqnarray}}
\newc{\eea}{\end{eqnarray}}
\newc{\ol}{\overline}
\newc{\wt}{\widetilde}
\newc{\bs}{\boldsymbol}
\newc{\m}{\mathcal}
\newc{\ra}{\rightarrow}
\newc{\lra}{\leftrightarrow}
\newc{\ba}{\begin{eqnarray}}
\newc{\ea}{\end{eqnarray}}
\newc{\pa}{\partial}
\newc{\vt}\vartheta
\newc{\D}{\Delta}
\def\b{\beta}

\newc{\nn}{\nonumber}

\def\beq{\begin{equation}}
\def\eeq{\end{equation}}
\def\bea{\begin{eqnarray}}
\def\eea{\end{eqnarray}}

\def\b{\textcolor{blue}}

\newcommand{\ov}{\overline}

\begin{document}

\begin{titlepage}

\vspace*{0.7cm}

\begin{center}
{
\bf\LARGE
 Aspects of  {\cal F}-Theory GUTs}
\\[12mm]
George~K.~Leontaris
\footnote{ \texttt{Based on Talk presented at Corfu 2011 Workshop  "Fields and Strings:
Theory - Cosmology - Phenomenology. September 14 - 18, 2011". (Prepared for the proceedings)}}
\\[-2mm]

\end{center}
\vspace*{0.50cm}
\centerline{ \it
Physics Department, Theory Division, Ioannina University,}
\centerline{\it GR-45110 Ioannina, Greece}
\vspace*{1.20cm}

\begin{abstract}
\noindent
The basic tools for model building in F-theory are reviewed  and applied to the construction of
 $SU(5)$ models. The flux mechanism  for gauge symmetry breaking  and doublet triplet splitting is analysed.
 A short account for the gauge coupling unification and the role of flux and Kaluza-Klein thresholds is
 also given.
 \end{abstract}

 \end{titlepage}

\thispagestyle{empty}
\vfill
\newpage

\setcounter{page}{1}

 \section{Introduction}

During the last four decades  there has been significant progress in our understanding of the world of elementary particles. The predictions of Standard Model (SM) of Electroweak and Strong interactions developed in early 70's  are now confirmed by an enormous amount of experimental data.  Nowadays,  highly sophisticated experiments like those performed at the Large Hadron Collider (LHC) at CERN intend to find the missing ingredient -the Higgs field- to complete the anticipated picture of the SM. However, from the theoretical point of view,  it was soon realised that the SM falls rather short for a complete and final theory of elementary particles and their fundamental forces.  Among other shortcomings, the SM involves a large number of arbitrary parameters,  while the gauge symmetry of the model is a product of gauge groups rather than a simple unified one.  Furthermore, gravity is not included, therefore the SM cannot be considered a truly unified theory of all fundamental forces.  However, from the extrapolation of the three SM gauge couplings there are hints indicating that they probably merge at a common mass scale at high energies and therefore it is expected that indeed, there is a larger symmetry, i.e., a Grand Unified Group (for reviews see~\cite{Langacker:1980js})  with the SM gauge group incorporated in it. Nevertheless, the SM alone cannot account for this since it is plagued by quadratic divergences at large energies where the Unification of couplings is expected.  The large difference between the weak energy scale where the strong,  and electroweak interactions are manifest (of the order of 100 GeV) and the expected energy where they unify  (around $10^{16}$ GeV),  is known as the hierarchy problem. There is belief that these difficulties might be evaded if supersymmetry is introduced.  There are significant theoretical reasons  indicating that if supersymmetry (for reviews see~\cite{Nilles:1983ge}) is indeed the solution to the SM drawbacks,  then it should be relevant at low scales accessible to present day experiments.  The existence of superpartners will be checked at LHC in the forthcoming years.

 Essential role in the theoretical developments and in particular the way supersymmetry is contributing to the solutions of the various theoretical issues,  has been played by String Theory according to which our world is ``immersed" in a ``hyperspace" consisting of ten space-time dimensions where six of them are compactified and extremely small to be observed. String Theory reconciles in a nice way supersymmetry, Grand Unified symmetries and unification of gauge couplings at a high (string) scale. Besides, quantization of gravity occurs naturally in the context of String Theory since the ultraviolent infinities can be avoided.
Thus, one of the most important tasks is to embed the successful Standard Model of electroweak and strong interactions in a unified String derived model. The unification of all interactions can be realised in a quantum gravity theory free of anomalies. At present, the only candidate theory for this role is String Theory.

 Some two  decades ago a major effort had been devoted to develop unified models in the context of Heterotic String Theory and there was a perception that it was the only one that includes the Standard Model. The great progress made in recent years has shown that other theories such as Type I Strings can also reproduce the Standard Model.
One of the interesting features of Type I string theory is that the scale where the theory leaves its trace
could in principle be very low~\cite{Antoniadis:1990ew}, even at the order of a few TeV, and therefore gives us the opportunity to solve the problem of hierarchy without requiring the existence of supersymmetry.
 Additionally the low unification scenario allows the possibility to seek experimental evidence  in  appropriately designed experiments. In this scenario an important role is played by extensive solitonic-type objects that appear in the Type I theory and are known as Dp-branes~\cite{Polchinski:1996na}. Our world could be localised on such a  brane immersed in a higher dimensional space. In this scenario, known as Brane-World scenario, the interactions of the Standard Model are confined on the brane while the gravitational interactions are spread throughout the
 whole 10d space and this explains the fact that the gravitational interactions in four-dimensional world are weaker compared to other fundamental interactions.

The last decade,  considerable efforts were concentrated in model building and the fermion masses from intersecting D-brane configurations (for related reviews see~\cite{Blumenhagen:2005mu}) embedded in a ten dimensional space.   In effective field theory models emerging from  intersecting D-branes, the matter fields are represented by strings attached on pairs of different D-brane stacks and they are localised at the intersections.
The gauge symmetry of these constructions consists of gauge group factors $U(n_1)\times\cdots\times U(n_k)$,  with matter accommodated in the various available  bifundamental  representations. Hence, in this context, the Standard Model gauge group could naturally emerge from some appropriate D-brane configuration. Since the various D-brane stacks span different dimensions of the ten dimensional  space, the corresponding gauge couplings $g_{1,\dots,k}$,  depend on different world volumes and therefore they generally have different values. Therefore, although there are many interesting features and success in the above approach, these models do not incorporate the anticipated gauge coupling unification in a natural way since there is no underlying symmetry that would force these couplings to be equal at the unification (string) scale. We note that it is possible  to assume  a D-brane set up with  $U(5)$ gauge group\footnote{Notice that in intersecting D-brane constructions of this type, the available gauge symmetries are of $U(N)$ and $SO(N)$ type, whilst
exceptional groups are absent.}, which contains all SM group factors in a single gauge symmetry and leads automatically to gauge coupling unification at the string scale. The main shortcoming of this possibility however, -in the context
 of intersecting D branes- is the absence of the tree level perturbative Yukawa coupling $10_M\cdot 10_M\cdot 5_H$  to provide fermion masses. We will see how these issues are resolved in the context of F-theory models.

\section{The Framework}

Work done during the last few  years provides convincing evidence that the above drawbacks can be evaded
when the desired grand unified theory symmetries (GUTs) are realised in F-theory\cite{Vafa:1996xn}\footnote{For
 reviews see~\cite{Denef:2008wq,Weigand:2010wm,Heckman:2010bq}.} compactified on Calabi-Yau fourfolds.
Recent progress in F-theory model building~\cite{Donagi:2008ca}-\cite{Heckman:2009mn} has shown that old successful GUTs including the  SU(5),  SO(10) models etc, are naturally realised on the world-volume of non-perturbative seven branes wrapping appropriate compact surfaces. The rather interesting fact in F-theory constructions is that because they are defined on a compact elliptically fibered Calabi-Yau complex four dimensional Manifold the exceptional groups ${\cal E}_{6,7,8}$, can be naturally incorporated into the theory too~\cite{Donagi:2008ca,Beasley:2008dc,Beasley:2008kw,Heckman:2009mn}.
  Although exceptional gauge symmetries suffer from several drawbacks when realised in the context of four-dimensional grand unified theories, in the case of F-theory models they are more promising as new possibilities arise for the symmetry breaking mechanisms and the derivation of the desired massless spectrum.

Present studies  have  led to remarkable progress on model building in F-theory~\cite{Heckman:2008qa}-\cite{Grimm:2011tb} with  a considerable amount of them  focusing  on three generation  $SU(5)$-GUT models. The vital issues of proton decay, the Higgs mixing term and the fermion mass structure require the  computation of Yukawa couplings~\cite{Heckman:2008qa,Font:2008id,Conlon:2009qq,Dudas:2009hu,King:2010mq,Dudas:2010zb,Leontaris:2010zd,Cecotti:2009zf,Ludeling:2011en,Krippendorf:2010hj,Aparicio:2011jx}.
F-model building gave rise to some interesting mechanisms to generate Yukawa hierarchy either
  with the use of fluxes~\cite{Heckman:2008qa,Cecotti:2009zf}  and the notion of T-branes~\cite{Cecotti:2010bp}  or with the implementation of the Froggatt-Nielsen mechanism~\cite{Dudas:2009hu,King:2010mq,Dudas:2010zb,Leontaris:2010zd,Ludeling:2011en}.
In~\cite{Cecotti:2009zf} (and further in~\cite{Aparicio:2011jx,Marchesano:2009rz}) it is shown that when three-form fluxes are turned on in F-theory compactifications, rank-one fermion mass matrices receive corrections, leading to
masses for lighter generations and CKM mixing.
 Flipped $SU(5)$~\cite{Heckman:2008qa,Jiang:2009za,King:2010mq, Kuflik:2010dg,Chen:2010tp,Chung:2010bn},
as well as some examples of $SO(10)$ F-theory models~\cite{Heckman:2008qa,Chen:2009me,Chen:2010ts}
 were also considered. Many  of these models predict exotic states below the unification scale,
and the renormalization group (RG) analysis of gauge coupling unification including the effect of such states and flux effects has been discussed in a series of papers
\cite{Blumenhagen:2008aw}-\cite{Leontaris:2011tw,Dolan:2011aq}. Other phenomenological issues such as
neutrinos from KK-modes\cite{Bouchard:2009bu},
proton decay \cite{Grimm:2010ez} and the origin of CP violation~\cite{Heckman:2009de} have also been discussed.
A systematic  classification of  semi-local F-theory GUTs arising from
a single ${\cal E}_8$ point of local enhancement, leading to simple GUT groups
based on ${\cal E}_6$, $SO(10)$ and $SU(5)$ on the del Pezzo surface has been presented in~\cite{Callaghan:2011jj}.
Here I focus on some phenomenological aspects of effective F-theory  models mainly with $SU(5)$ symmetry.
To make this presentation self-contained in the next section I review in brief the basics of F-theory
and elliptic fibration. Section 4 is devoted to the methodology of F-theory model building. In section
5 the spectral cover approach is reviewed whilst the remaining sections deal with various phenomenological
issues of specific examples in the context of $SU(5)$ models.

\section{Rudiments of F-theory and Elliptic fibration}

We start with a short description of the salient features of
F-theory and F-theory model building following mainly the works
of~\cite{Donagi:2008ca} and \cite{Beasley:2008dc,Beasley:2008kw}\footnote{see also
  \cite{Heckman:2010bq,Braun:2010ff}}.
 F-theory can be considered as a 12-dimensional theory  which arises
from the geometrization of the  type IIB 10-dimensional string theory.
The effective theory is described by the  type IIB supergravity whose
 bosonic field content contains  the metric $g_{MN}$ the dilaton field $e^{\phi}$ and
 the $p$-form potentials $C_p,\,p=0,2,4$ which imply the corresponding field strengths $F_{p+1}=d\,C_p$.
 An important observation is that when $p$-form magnetic fluxes are turned on
 in the internal manifold, new string vacua may appear and a tree-level
 moduli potential will be generated. Further, there are two scalars contained in
 the aforementioned bosonic spectrum, namely $C_0$ and $e^{\phi}$
  which  can be combined into a complex  modulus
 \ba
 \tau =C_0+\imath\,e^{-\phi}\equiv C_0+\frac{\imath}{g_s}\label{modulus}
 \ea
  In addition,  it is convenient to define the field combinations
 \ba
 G_3&=&F_3-\tau H_3\\
\tilde F_5&=&F_5-\frac 12 G_2\wedge H_3-\frac 12 B_2\wedge F_3\label{5F}
\ea
The 5-form field defined in (\ref{5F}) has to obey the selfduality condition
$\ast\tilde F_5=\tilde F_5$ where $\ast$ stands for the Hodge star. With these
ingredients one can write an action leading to the correct equations of
motion~\cite{Denef:2008wq}
\[
S_{IIB}\propto
 \int d^{10}x\sqrt{-g}\,R-\frac 12\int \frac{1}{({\rm Im}\tau)^2}d\tau\wedge\ast
 d\bar\tau+\frac{1}{{\rm Im}\tau}G_3\wedge \ast\ov{G}_3+\frac 12
\tilde F_5\wedge \ast\tilde F_5+C_4\wedge H_3\wedge F_3\nn
\]
The action is invariant under the following $SL(2,Z)$ duality transformations
\ba
\tau&\ra&\frac{a\tau+b}{c\tau+d}\\
\left(\begin{array}{c}H\\F\end{array}\right)&\ra&\left(\begin{array}{cc}d&c\\b&a\end{array}\right)
\left(\begin{array}{c}H\\F\end{array}\right)
\ea
together with $\tilde F_5\ra \tilde F_5$ and $g_{MN}\ra g_{MN}$.
This action looks like it has been obtained from a compactified 12-dimensional theory on a torus
with modulus $\tau$ defined in (\ref{modulus}).
The $F_3,H_3$ fields appear in $S_{IIB}$ as if they have been obtained from a 12-d
field strength $\widehat F_4$ reduced along the two radii of the torus. In
F-theory  $\tau$ is interpreted as the complex structure modulus of an elliptic curve generating
a complex fourfold which constitutes the elliptic fibration over the CY threefold. Since
the fibration relies on the $\tau=C_0+\imath/g_s$, this means that the gauge coupling is
not a constant and the resulting compactification is not perturbative.  Hence,
according to the above picture,  F-theory\cite{Vafa:1996xn} is defined on a background
$R^{3,1}\times X$  with $R^{3,1}$ our usual space-time and $X$ an elliptically
fibered Calabi-Yau (CY)  complex fourfold with a section over a complex three-fold base $B_3$.

 In~\cite{Sen:1996vd} a specific example was presented  where
   there is an equivalence between F-theory compactifications on a K3 surface
 and the Heterotic theory compactification on $T^2$.
 A K3 surface is a complex smooth regular manifold with trivial canonical bundle.
 The general elliptically fibered K3 is described by
the Weierstrass equation
   \ba
 { y^2}&=&{ x^3}+{ f(z,w)}\,{ x}\,u^4+{g(z,w)}\,u^6\label{Wei0}
   \ea
where ${z,w,x,y,u}$ are parameters  of the fibration and  ${f,g}$  homogeneous polynomials
of degree 8 and 12 respectively. The equation is invariant under the following
two rescalings
\[\{z,w,x,y,u\}\ra  \{\lambda z, \lambda w, \lambda^4x,\lambda^6 y,u\}\;;\;
\{z,w,x,y,u\}\ra  \{z,  w, \mu^2x,\mu^3 y,\mu u\}\]
Indeed, for the first rescaling the left hand side becomes $y^2\ra \lambda^{12} y^2$
 and the same weight  emerges for the right hand side of (\ref{Wei0}).
 Similarly,  one
finds that for the second rescaling from both terms of the equation a weight $\mu^6$
is factored out.
There are five coordinates compared to two rescalings and one equation, thus
the equation describes a two complex dimensional surface.
For the first  rescaling  we observe that the sum of the weights
is $1+1+4+6+0=12$, i.e. equal to the weight 12, and the second is $0+0+2+3+1=6$ is
equal to the weight of the second equivalent equation. Therefore, this is a CY manifold.

Fixing $u=1,w=1$ the above equation becomes
\ba
y^2&=&x^3+f(z)x+g(z)\label{Wei1}
\ea
We now observe that $f,g$  transform as sections $f\in K_{B_3}^{-4}, g\in K_{B_3}^{-6}$.  This can be understood if we assign the scalings $x\ra \lambda^2 x$ and $y\ra \lambda^3 y$ so that (\ref{Wei1}) becomes $\lambda^6\,y^2=\lambda^6 x^3+\tilde f\,\lambda^2\,x+\tilde g$ implying $\tilde f\ra \lambda^4\,f$ and $\tilde g\ra \lambda^6\,g$.

\begin{figure}[h]
\centering
\includegraphics[scale=.4]{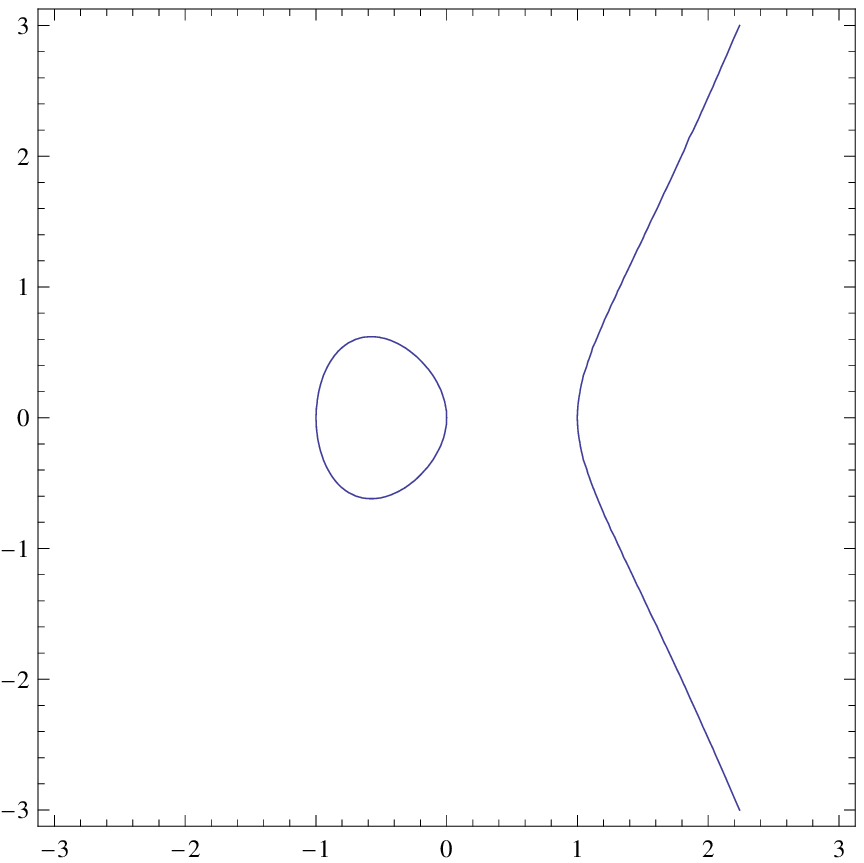}
\includegraphics[scale=.4]{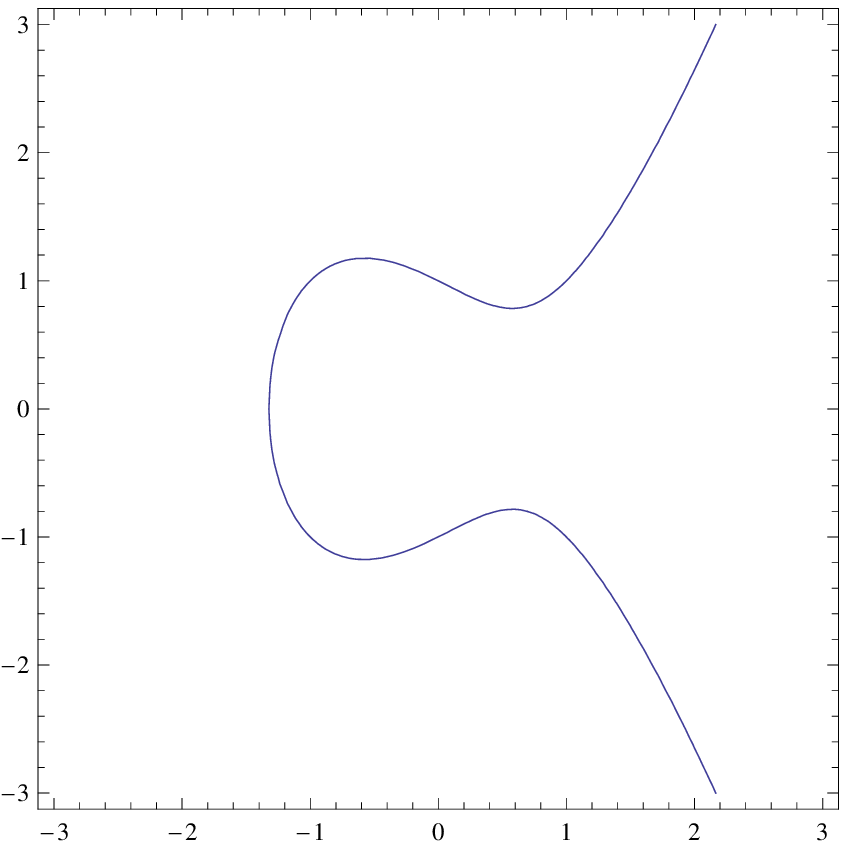}
\includegraphics[scale=.4]{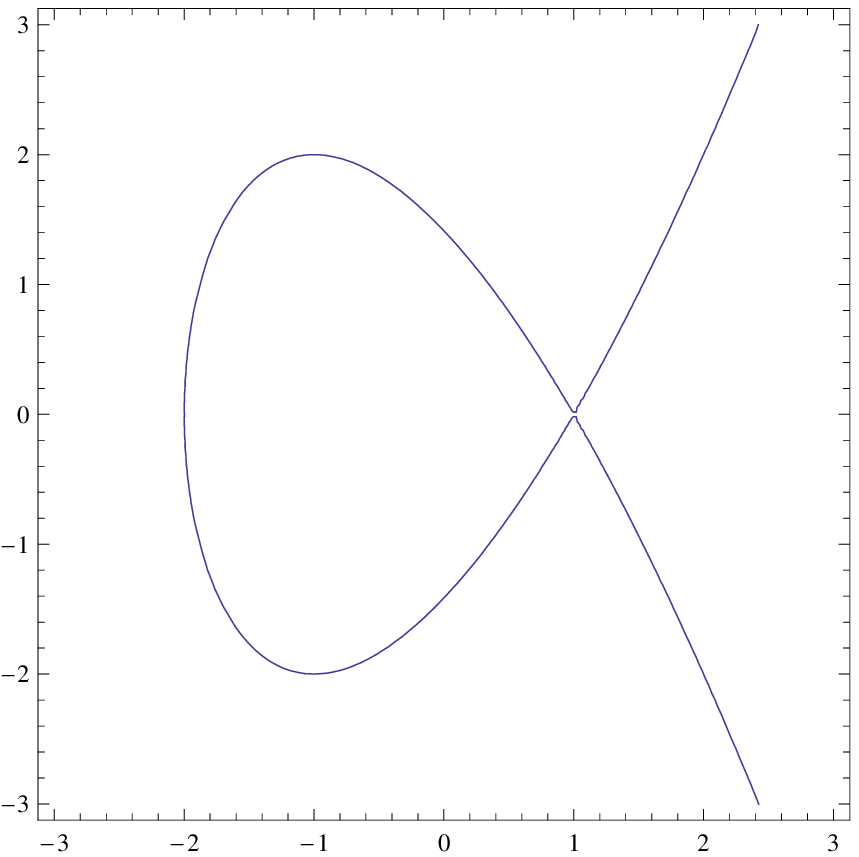}
\caption{Fixing the values of the polynomials (f,g) to certain
 real numbers in the Weierstra{\ss} equation, elliptic fibrations reduce to elliptic curves.
 The three cases correspond to the three possible cases of the discriminant,
 bigger, smaller or equal to zero respectively. }
 \label{xxx}
\end{figure}

The functions  $f(z), g(z)$ now are considered 8 and 12 degree polynomials in $z$.
For each point of the base, the equation describes a torus labeled by the coordinate $z$.
(To get an intuition, note that fixing $f,g$ to be real numbers, (\ref{Wei0}) reduces to elliptic curves, see fig.\ref{xxx}).

The modular parameter of the torus is related to the functions $f,g$ through the $SL(2,Z)$
modular invariant function $j(\tau)$
\ba
j(\tau)&=&\frac{4 (24 f)^3}{4{ f}^3+27 { g}^2}\label{modf}
\ea
where
\ba
j(\tau)&=&e^{-2\pi i\tau}+744+{\cal O}(e^{2\pi i\tau})
\ea
The curve described by (\ref{Wei0}) is non-singular provided that the discriminant
\ba
{ \Delta} &=& 4\,{ f}^3+27\, { g}^2\label{Discr}
\ea
in non-zero.  At the zero loci of the discriminant $\Delta$, (i.e., at $ { \Delta} =0$ )
the elliptic curve becomes singular with one cycle shrinking to zero size and
the fiber degenerates\footnote{The Discriminant locus may have several irreducible components, so that
$\Delta =\sum_in_i S_i$
where $S_i$ are the divisors of $B_3$ and $n_i$ represent their multiplicities. The singularities of
the CY 4-fold are developed along the divisors with $n_i>1$.}.
There are 24 zeros $z_i$ of the discriminant which are in general distinct and different from
the zeros of $f,g$.  This corresponds to 24 7-branes located at $z_i,\,i=1,2,\dots,24$.
In the vicinity of such a point using (\ref{modf}) and (\ref{Discr}) we have
\ba
j(\tau(z))\sim\frac{1}{z-z_i}&\ra&\tau(z)\approx \frac{1}{2\pi i}\log(z-z_i)
\ea
up to $SL(2,Z)$ transformations. In the limit $z\ra z_i$, we observe that $\tau\ra i\,\infty$
and since $\tau=C_0+i/g$ this means that we are in the weak coupling regime since $g\ra 0$. Further,
since $\ln(z-z_i)=\ln|z-z_i|+i\,\theta$, performing a complete rotation around $z_i$, $\tau$
undergoes a monodromy $\tau\ra \tau+1$, or equivalently
\[ C_0\ra C_0+1,\ra \oint_{z_i}F_1=\oint_{z_i}dC_0=1\]
This implies the existence of a 7-brane at $z_i$, while totally there are 24 such branes in the
compact transverse space. However, since the space is compact the sum $\oint_{z_i}F_1$ must vanish.
Further considerations along these lines lead to the conclusion that F-theory is strongly coupled.
There are limiting cases however where F-theory has a perturbative expansion. Indeed, suppose
that $f^3/g^2$ is constant which can be satisfied by assuming that~\cite{Sen:1996vd}
\[g=\phi^3,\; f=a\phi^2,\; \phi=\prod_{i=1}^4 (z-z_i)\]
Substituting one finds
\[j(\tau)=\frac{4 (27a)^3}{27+4a^3}\]
which gives a weak coupling regime everywhere on the base for $27+4a^3\approx 0$ i.e., $a\sim -\frac{3}{4^{1/3}}$.

We discuss now how this geometric picture is associated to the gauge group structure
and the spectrum of an effective low energy theory model.
Recall first that in intersecting  D-brane constructions non-abelian gauge symmetries emerge when
more than one D-branes coincide. While a single D-brane is associated to a $U(1)$
symmetry, when $n$ of $D6$ branes coincide the gauge group becomes
$SU(n)$.  In F-theory when D7 branes coincide at certain point, then at this point there is
a singularity of the elliptic fibration.  The  {\it singularities} of the manifold are  classified with respect to the vanishing
order of the polynomials ${ f,g}$ and the zeros of the discriminant $\Delta$. They  determine the {\it gauge group}
and {\it the matter content} of the {\it F-theory} compactification.  By adjusting the coefficients of the
polynomials $f,g$ we can obtain ${\cal A}, {\cal D}, {\cal E}$ types of gauge groups.

\begin{figure}[h]
\centering
\includegraphics[scale=.6]{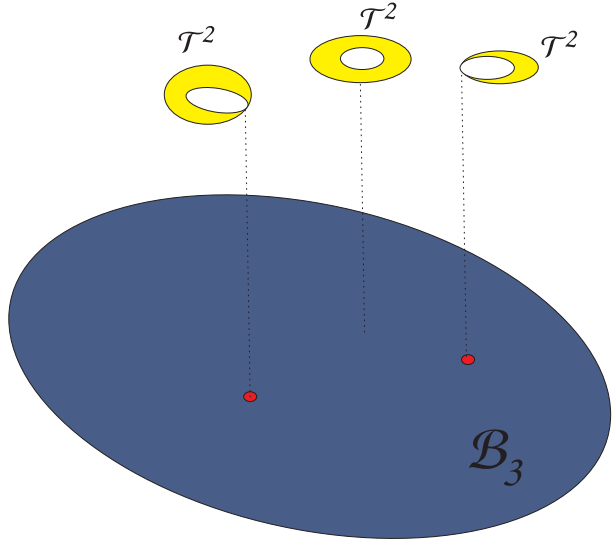}
\caption{{CY four-fold constituting an elliptic fibration over a three-fold base $B_3$ (only two dimensions are shown).
Every point of $B_3$ is represented by a torus with modulus $\tau=C_0+\imath/g_s$. Red points represent 7-branes, orthogonal to $B_3$. The torus degenerates at
these `points' (vanishing cycle). Going around the non-trivial cycle, the vanishing cycle
undergoes monodromy.  }}
\end{figure}

\subsection{Tate's Algorithm}

According to the interpretation above,  in F-theory
 the  gauge symmetry is associated to the singularities of the internal
compact manifold.  A systematic  analysis of these singularities
has started with the work of Kodaira~\cite{Kodaira}.  Given the form of the
Weierstrass equation
\[ y^2=x^3+f(z)x+g(z)\]
the Kodaira classification relies on the vanishing order of the
polynomials $f,g$ and the discriminant $\Delta$.  This is summarised
in Table~\ref{T_K}.  A useful tool for the analysis of the gauge properties
of an F-theory GUT is Tate's algorithm~\cite{Tate1975}~\footnote{For recent advances
on Tate's classification see~\cite{Katz:2011qp,Morrison:2011mb,Esole:2011sm}.}.
Tate's Algorithm gives an algorithmic procedure to describe  the singularities
of the elliptic fiber and determine the local properties of the associated  gauge
group.

We follow~\cite{Katz:2011qp} to
sketch how this analysis works for a few simple cases. We assume a small set
$U$ of the base  and that the restriction $S|_U$ will have a defining equation
$\{z=0\}$. In this patch we expand the coefficients of the Weierstrass
equation in powers of  $z$
\begin{equation}
f(z)=\sum_{n}f_nz^n,\;
g(z)=\sum_{m}g_mz^m
\label{beta2}
\end{equation}
\begin{table}
\begin{center}
\begin{tabular}{|l|l|l|l|l|}
ord($f$)& ord($g$)& ord($\Delta$) &fiber type &Singularity\\
\hline
$0$&$0$&$n$&$I_n$&$A_{n-1}$
\\
$\ge 1$&$1$&$2$&$II$&none
\\
$1$&$\ge 2$&$3$&$III$&$A_1$
\\
$\ge 2$&$2$&$4$&$IV$&$A_2$
\\
2&$\ge 3$&$n+6$&$I_n^*$&$D_{n+4}$
\\
$\ge 2$&$ 3$&$n+6$&$I_n^*$&$D_{n+4}$
\\
$\ge 3$&4&8&$IV^*$&${\cal E}_6$
\\
$ 3$&$\ge 5$&9&$III^*$&${\cal E}_7$
\\
$\ge 4$&5&10&$II^*$&${\cal E}_8$
\\
\hline
\end{tabular}
\caption{Kodaira's classification of Elliptic Singularities}
\end{center}
\label{T_K}
\end{table}
Plugging  into the discriminant the above expansions  we get
\begin{equation}
\label{Disc}
\begin{split}
\Delta &=4\left[f_0+f_1z+ O(z^2)\right]^3+27\left[g_0+g_1z+ O(z^2)\right]^2\nn
\\
      &=4f_0^3+27g_0^2+(12 f_1 f_0^2+54 g_0 g_1)z+(12 f_2 f_0^2+12 f_1^2 f_0+27 g_1^2+54 g_0 g_2)z^2+O(z^3)
\end{split}
\end{equation}
We can demand  $z/\Delta$ which requires that  the zeroth order coefficient is zero, i.e.
$4f_0^3+27g_0^2=0$. Assuming that  $f_0,g_0$ are simple functions of a new variable $t$,
$f_0=a t^2, g_0=b t^3$, the coefficients $a,b$ must obey
\[ 4 a^3+27 b^2=0 \]
which is satisfied for $a=-1/3,b=2/27$, thus
\begin{equation}
f_0=-\frac 13 \,\,t^2,\;\;
g_0=\frac{2}{27}t^3
\label{fgt}
\end{equation}
The discriminant now becomes
\begin{equation}
\begin{split}
\label{Del1}
\Delta &= \frac 43 t^3\left( f_1 t+ {3}g_1\right)z+\left( \frac{4 f_2 t^4}{3}+4 g_2 t^3-4 f_1^2 t^2+27 g_1^2\right)z^2+O(z^3)
\end{split}
\end{equation}
We turn now to the Weierstrass equation. To put it in the Tate form we make the substitution
\[ x\ra x+\frac 13\,t\]
Substituting and reorganising in powers of $x$, we get
\[ y^2= x^3+t\,x^2 +(f_2 z^2+f_1 z)\,x+\frac 13\left[ ( f_1t+3 g_1)z+( f_2t+3 g_2 ) z^2+( f_3t+3 g_3 ) z^2+\cdots\right]\]
By redefining $g_i\ra \tilde g=g_i+f_it/3$ to absorb the terms $\sim t$, we write
\[ y^2= x^3+t\,x^2 +(f_1 z+f_2 z^2+\cdots)\,x+ \tilde g_1  z+  \tilde g_2z^2+\cdots\]
This is the Tate form $I_1$. The discriminant takes also the simpler form
\begin{equation}
\begin{split}
\label{Del2}
\Delta &= 4 t^3\tilde g_1\,z+\left( 4 \tilde{g}_2 t^3-f_1 \left(18 \tilde{g}_1+t f_1\right) t+27 \tilde{g}_1^2\right)z^2+O(z^3)
\end{split}
\end{equation}

Let us now examine the conditions to obtain $z^2/\Delta$. From the form (\ref{Del2})
obtained for $\Delta$
we see that the coefficient  of  $z$ is zero if
\[\tilde g_1\equiv \frac{ f_1 }{3}t+ g_1=0\]
This condition eliminates also several other terms, the result being
\[\Delta = (4 \tilde{g}_2 t-f_1^2)t^2\,z^2+O(z^3)\]
In addition, the Weierstrass equation becomes
\[ y^2= x^3+t\,x^2 +(f_1 z+f_2 z^2+\cdots)\,x+   \tilde g_2z^2+\cdots\]
which  is the Tate form for $I_2$ in Table~\ref{dpn}. For global obstructions
with regard to the general validity of the Tate forms see~\cite{Katz:2011qp}.

The procedure can be continued for the next order and so on. Partial results
are summarised in the Table~\ref{dpn}. (for complete results see Table of
refs\cite{Tate1975, Bershadsky:1996nh,Katz:2011qp}).

We then write the  general Tate  form of the Weierstrass equation as follows
   \ba
 {y^2}+{ a_1}{ x\,y}+{ a_3}{ y}&=&
 { x^3}+{a_2}\,{ x^2}+{ a_4}{ x}+{ a_6}\label{Wei}
   \ea
   with $a_n$ being polynomial functions on the base.
The indices of the coefficients $a_n$ have been chosen so to indicate the section they
belong to, i.e. $a_n\in K_{B_3}^n$. Thus  each term is a section $K^{-6}_{B_3}$.

To make connection with the  previous standard form (\ref{Wei1}) of the Weierstrass equation
 we complete  the square and the cube as follows. We form the square on the left hand side
\ba
\left(y+\frac{a_1x+a_3}{2}\right)^2&=& { x^3}+{a_2}\,{ x^2}+{ a_4}{ x}+{
a_6}+\left(\frac{a_1x+a_3}{2}\right)^2\nn
\ea
while we equate the RHS with
\[\left(x+\lambda\right)^3+f\,(x+\lambda)+g\]
Comparing, we get
\ba
f&=&\frac{1}{48} \left(24 \,a_1 \,a_3-\left(a_1^2+4 \,a_2\right)^2\right)+ a_4\nn\\
g&=&\frac{1}{864} \left(a_1^6+12 a_1^4 a_2-36 a_1^3 a_3+48 a_1^2 a_2^2\right.\nn\\&&\left.
-72 a_4 \left(a_1^2+4
   a_2\right)-144 a_1 a_2 a_3+64 a_2^3+216 a_3^2\right)+a_6
\ea
Using the definitions
\[
\beta_2=a_1^2+4 a_2,\;\beta_4=a_1a_3+2a_4,\;\beta_6=a_3^2+4 a_6\]
the functions $f,g$ can be rewritten in a simpler form
\ba
f&=&-\frac{1}{48}\left(\beta_2^2-24 \beta_4\right)\nn\\
g&=&-\frac{1}{864}\left(-\beta_2^3+36\beta_2\beta_4-216\beta_6\right)
\ea
If we further define
\[\beta_8=\beta_2a_6-a_1a_3a_4+a_2a_3^2-a_4^2\]
we can write the discriminant
\[\Delta=-\beta_2^2\beta_8-8\beta_4^3-27\beta_6^2+9\beta_2\beta_4\beta_6\]
$f,g$ are assumed to be functions of a complex coordinate $z$ on the base $B_3$.

In summary, we have the following  picture: assuming a hypersurface   ${S}\in { B_3}$
 singularity of ${\cal ADE}$ type at $z=0$ at a certain point of the base  we generate a  fibration
 of this base parametrised by the coordinate $z$. As $z\ne 0$ the original symmetry
 breaks leading to a subgroup. Going around this $z$-point
where the fiber degenerates we return to the same singularity up to a monodromy action.
In general the effect of the  monodromy action cannot be
absorbed by some gauge transformation and as a result the gauge symmetry is not fully
restored. Thus, one ends up with a reduced gauge symmetry.
The order of vanishing of ${a_i=b_i\,z^{n_i}}$ characterises the type of
 singularity, i.e, the {\bf gauge group} supported by the divisor ${ S}$.
For example, the choice
\[a_1=-b_5,a_2=b_4{ z},a_3=-b_3{ z^2},a_4=b_2{ z^3},a_6={ z^5}b_0\]
where $b_i$ are independent of $z$, lead to the equation
 \ba
 y^2=x^3+b_0 { z^5}+b_2 x { z^3}+b_3yz^2+b_4x^2 { z}+b_5xy\label{su5sing}
 \ea
which as can be seen from table \ref{dpn} implies an ${\bf SU(5)}$ Singularity. The coefficients $b_i$ are in general non-vanishing and can be seen as sections of line-bundles on $S$.
\begin{table}[!t]
\centering
\renewcommand{\arraystretch}{1.2}
\begin{tabular}{|c|c|c|c|c|c|c|c|}
\hline
Type &{\bf Group} & ${ a_1 }$& ${ a_2 }$& ${ a_3} $& ${ a_4 }$& ${a_6} $& ${ \Delta}$\\
\hline
$I_0$& ${0}$&0&0&$0$&$0$&$0$&$0$   \\
\hline
$I_1$& $-$&0&0&$1$&$1$&$1$&$1$   \\
 \hline
$I_2$&  $-$&0&0&$1$&$1$&$2$&$2$   \\
 \hline
$I_{2n}^s$& ${ SU(2n)}$&0&1&$n$&$n$&$2n$&$2n$   \\
\hline
$I_{2n+1}^s$& ${ SU(2n+1)}$&0&1&$n$&$n+1$&$2n+1$&$2n+1$   \\
 \hline
$I_1^{*s}$& ${ SO(10)}$&1&1&$2$&$3$&$5$&$7$   \\
 \hline
$IV^{*s}$& ${{\cal E}_6}$&1&2&$3$&$3$&$5$&$8$   \\
 \hline
$III^{*s}$& ${ {\cal E}_7}$&1&2&$3$&$3$&$5$&$9$   \\
 \hline
$II^{s}$& ${ {\cal E}_8}$&1&2&$3$&$4$&$5$&$10$  \\
 \hline
\end{tabular}
 \caption{Particular cases of  Tate's algorithm. (For the complete results see~\cite{Tate1975,Bershadsky:1996nh}.)
  The order of vanishing of the coefficients ${ a_i\sim z^{n_i}}$, the discriminant $\Delta$
   and the corresponding {gauge group}. The highest singularity allowed in the elliptic fibration is
   ${ {\cal E}_8}$.\label{dpn} }
\end{table}
We denote with  $c_1$ the $1^{st}$ Chern class of the {\bf Tangent} Bundle to $S_{GUT}$
and $-t$ the $1^{st}$  Chern class of the {\bf Normal} Bundle to  $S_{GUT}$.
It is also customary to define the quantity
\[\eta =6\,c_1-t\]
while $c_1(B_3)|_S=c_1(S)-t$.
Returning to (\ref{su5sing}) defining  the $SU(5)$ singularity, the various coefficients
$b_k$ and parameters $x,y,z$ are sections of line bundles as they appear in the following Table
\ba
\begin{array}{ll}
section&c_1(bundle)\\
x:&2(c_1-t)\\
y:&3(c_1-t)\\
z:&-t\\
b_k:&\eta-k\,c_1=(6-k)c_1-t
\end{array}
\ea
With these definitions, each term of the equation (\ref{su5sing}) is a section of the same class $6(c_1-t)$, in accordance with (\ref{Wei0}). Indeed, for example
\ba
b_2xz^3&:& \{(6-2)c_1-t\}+\{2c_1-2t\}-3t\;=\;6(c_1-t)
\ea
Substituting $a_i=b_iz^{n_i}$, the $\beta_k$ take the form
\ba
\beta_2&=&b_5^2+4b_4z\nn\\
\beta_4&=&b_3b_5z^2+2 b_2z^3\nn\\
\beta_6&=&b_3^2z^4+4b_0z^5\nn\\
\beta_8=\frac{\beta_2\beta_6-\beta_4^2}4&=&z^5({\cal R}+z(4b_0b_4-b_2^2))\nn\\
{\cal R}&=&b_3^2b_4-b_2b_3b_5+b_0b_5^2\nn
\ea
We can check how the symmetry is enhanced for certain choices. For example,
choosing $b_5=0$ we see that $\beta_2\propto zb_4,\,\beta_4\propto z^3b_2$ etc while
the discriminant becomes $\Delta\propto z^7$. Comparing with Tate's results in Table~\ref{dpn},
we see that this corresponds to an $SO(10)$ singularity. Thus, a matter curve is defined
 along the intersection with another brane where we expect to find the ${\bf 10}$
of $SU(5)$ in the adjoint decomposition of $SO(10)$, therefore  we write
\[\Sigma_{10}=\{b_5=0\}\]
Demanding ${\cal R}=0$, we see that $\Delta\sim z^6$ and this corresponds to an $SU(6)$
singularity. The $SU(6)$ adjoint induces the {$\bf 5$} of $SU(5)$, therefore we define
the matter curve
\[\Sigma_{5}=\{{\cal R}=0\}\]

Further enhancements are obtained setting additional coefficients equal to zero. They
result to triple intersections of branes which define points in the internal
geometry where the Yukawa couplings are formed. Choosing $b_4=b_5=0$, we can proceed
as above and see that we obtain an ${\cal E}_6$ enhancement. This involves the top Yukawa coupling.
Similarly, $b_3=b_5=0$ implies an $SO(12)$ enhancement which is the origin
of the bottom mass term:
\[\{b_5=b_4=0\}\ra\lambda_t,\;\{b_5=b_3=0\}\ra\lambda_b\]
Using the homology classes derived previously for $b_i$'s we can also deduce
those of the matter curves. In particular we find
\[[\Sigma_{10}]=c_1-t,\;[\Sigma_5]=8c_1-3t,\]
or
\[[\Sigma_5]-3[\Sigma_{10}]-5c_1=0\]
The last one is equivalent to the anomaly cancellation condition~\cite{Donagi:2009ra}.

\section{Model building}

In the previous section we have analysed in some detail  the geometric singularities
and their  interpretation as gauge symmetries. In the present section we describe
the basic steps for model building following closely the analysis of
\cite{Beasley:2008dc,Beasley:2008kw}.

The ultimate goal is to associate this geometrical conception to
a  GUT model and make a choice of a compact `surface' $S$ of suitable  topological type
to build an effective field theory with the desired massless spectrum.
Hence, the required  set up consists  of a 7-brane stack  wrapping a compact K\"ahler
surface $S$  of two complex dimensions while the gauge theory of a particular model is
 associated with the geometric singularity of
the internal space~\cite{Bershadsky:1996nh,Donagi:2008ca,Beasley:2008dc,Beasley:2008kw}.

To make  a more specific choice of $S$ we must require some further phenomenological
constraints. For example, we have pointed out in the introduction that the MSSM
spectrum drives the three SM gauge couplings to a common value at a high scale which is
nevertheless  at least two orders smaller than the Planck scale. It has been argued~\cite{Beasley:2008dc}
that in order to achieve  a decoupling limit of gravity
the spacetime filling sevenbrane  associated to the gauge symmetry $G_S$
must wrap a del Pezzo surface. The simplest ones are $\mathbb{P}^1\times \mathbb{P}^1$
and $\mathbb{P}^2$.  There are eight more  del Pezzo surfaces $dP_n$ constructed
from an operation known as `blow up' of $\mathbb{P}^2$ at generic points.
(for a detailed discussion see \cite{Beasley:2008dc,Beasley:2008kw}).
 We may further specify this choice   to the del Pezzo $dP_8$ surface
since all other del Pezzo surfaces can be obtained from this one by blowing down various
two cycles of the latter. In correspondence with the del Pezzo surfaces, it is now possible to
assume  singularities associated to exceptional gauge symmetries ${\cal E}_{8}$ and its
subgroups, which incorporate the known successful GUT symmetries such as $SU(5)$ and $SO(10)$.

We discuss now the breaking mechanism of the gauge group down to SM.
In general, in F-theory there are two mechanisms available.   Higgs mechanism
and fluxes (we mention also discrete Wilson lines used in the heterotic string,
but this mechanism will not be implemented here).  We aim to build a unified
theory, the minimal one being $SU(5)$, thus a Higgs breaking
 mechanism  requires the adjoint representation. But if $S$ is a del Pezzo surface,  zero mode adjoint
 Higgs fields are not at our disposal. Even if for some other choice of $S$ the Higgs adjoint is available,
 this usually leads to a conventional GUT model with resulting spectrum involving undesired matter fields.
For example, the $SU(5)$ GUT breaking by the  24-Higgs adjoint allows in the
spectrum the dangerous triplet fields which mediate proton decay. The alternative
possibility  is to turn on $U(1)$ fluxes on the worldvolume of the 7-brane.
We will see that the breaking of the GUT group with this mechanism gives the
opportunity to eliminate unwanted fields from the light spectrum.
We note in passing that in heterotic string theory the  $U(1)$ flux  mechanism
cannot be implemented  for the $SU(5)$ breaking. This would require
a flux along $U(1)_Y$ and the corresponding gauge boson
would develop a string scale mass via the Green-Schwarz
mechanism. On the contrary, in  F-theory  we can arrange so that although
 the cohomology class of the flux on the seven-brane can be non-trivial,
 it can  represent a trivial class in the base of the compactification.
Thus, we can break $SU(5)$ turning on a $U(1)_Y$ flux for example, while
keeping the corresponding gauge boson massless.

Next we come to the matter and Higgs fields.  Matter can be found in the bulk
from the decomposition of the adjoint representation as well as  on Riemann surfaces
which are located at the intersection between the GUT model seven-brane and additional
seven-branes. In several cases, the bulk matter can be of exotic type and has
to be eliminated by some suitable condition. We will see that this is possible
when the GUT symmetry is $SU(5)$ but it is not true for $SO(10)$ and possible higher
groups.  It is possible however to turn on  singlet vevs  with appropriate $U(1)$-`charges'
to make some of these states massive or to associate  some of them to ordinary matter.

Suppose then that we start with a legible gauge symmetry group $G_S$ associated
to the singularity of $S$.
To determine the massless spectrum  of a given GUT model, we first turn on a
non-trivial background field configuration on
$S$  along some subgroup $H_S$ of $G_S$.
Then the effective field theory gauge group is given by the commutant subgroup
of $H_S$ in $G_S$,  i.e,
\[G_S\supset\Gamma_S\times H_S\]

Let us start with the matter in the bulk.
 The spectrum is found in  representations which arise from the decomposition of
 the adjoint of $G_S$ under $\Gamma_S\times H_S$
\footnote{  If $H_S$ contains semi-simple $U(1)$ factors, $\Gamma_S$ corresponds to a proper
subgroup of the four-dimensional subgroup. This is the case of $G_S={\cal E}_6$
with $H_S$, where  the commutant is $SO(10)\times U(1)$.}
\ba
{\rm ad}(G_S)=\bigoplus \tau_j\otimes T_j
\ea
In general, the net number of chiral  minus anti-chiral
states is given in terms of a topological index
formula~\cite{Beasley:2008dc},
$ n_{\tau}-n_{\tau^*}=\chi(S,{\cal T}_j^*)-\chi(S,{\cal T}_j)$
where $\tau^*$ is the dual representation of $\tau$, ${\cal T}$ is the bundle transforming in the
representation $T$ and $\chi$ is the Euler character~\footnote{For
$H_S=SU(n)$, the spectrum on the bulk is always non-chiral
since the corresponding instanton solutions have vanishing first Chern class.}. If $h^i={\rm dim}_CH^i$,
i.e. the dimension of the Dolbeault cohomology groups, then $\chi=h^0-h^1+h^2$.
Moreover, if $S$ is a del Pezzo (or Hirzebruch) surface then $H^2_{\bar\partial}(S, T_j)=0$ while
 when the holomorphic bundle $T_j$ is irreducible and non-trivial we also have $H^0_{\bar\partial}(S, T_j)=0$.

As a more specific  example, let us assume that the bulk gauge group is ${\cal E}_6$.
Under the decomposition ${\cal E}_6\supset SO(10)\times U(1)$, we get
\ba 78&\ra& 45_0+1_0+16_{-3}+\overline{16}_3
\ea
Thus, in addition to the adjoint of $SO(10)$ we also get the
zero modes $16_{-3}$ and $\overline{16}_3$  characterised by the line bundles
$ {\cal L}^{-3}$ and ${\cal L}^{+3}$ respectively.
If we assign the number of states by $n_{16}$ and $n_{\overline{16}}$ respectively
in order to obtain chirality we need to have $n_{16}-n_{\overline{16}}\ne 0$.
We recall that the    number of states is minus the Euler character $n_{16}=-\chi(S,L)$.
For $S$ a del Pezzo and ${\cal L}$ a line bundle over $S$, the Riemann-Roch theorem states
\ba
\chi(S,{\cal L})&=&1+\frac 12 c_1({\cal L})\cdot  c_1({\cal L})+\frac 12 c_1({\cal L})\cdot c_1(S)\nn\\
\chi(S,{\cal L}^*)&=&1+\frac 12 c_1({\cal L}^*)\cdot  c_1({\cal L}^*)+\frac 12 c_1({\cal L}^*)\cdot c_1(S)\nn\\
&=&1+\frac 12 c_1({\cal L})\cdot  c_1({\cal L})-\frac 12 c_1({\cal L})\cdot c_1(S)
\ea
Therefore, the difference
\[\chi(S,{\cal L}^*)-\chi(S,{\cal L})=- c_1({\cal L})\cdot c_1(S)\]
counts the number of chiral states $n_{16}-n_{\overline{16}}$.

Now in this set up one may assume other 7-branes  spanning different directions
of the internal space. In particular, when
other  seven branes $S_1, S_2,\cdots$ intersect with the GUT brane wrapping the surface $S$, they form
 Riemann surfaces~\footnote{A Riemann Surface (RS) is a connected Hausdorff topological space
together with a complex structure; according to the Riemann famous mapping theorem, a simply
connected RS  is isomorphic to: the Riemann sphere, or to $\mathcal{C}$, or to the open unit disc
$|z|<1, z\in \mathcal{C}$.} denoted subsequently with $\Sigma_i$.
Chiral matter  and Higgs fields reside on  these Riemann surfaces thus  we call them matter curves.
Along these matter curves gauge symmetry is enhanced. These chiral states appear in bifundamental
 representations in close analogy to the case of intersecting D-brane models. Along the intersections the rank of the singularity increases. Designating $G_S$ the gauge group on the surface $S$ and $G_{S_i}$ that associated with  $S_i$, the gauge group on $\Sigma_i$ is enhanced to $G_{\Sigma_i}\supset G_S\times G_{S_i}$ whose   the adjoint in general  decomposes as
\ba
{\rm ad}(G_{\Sigma_i})= {\rm ad}(G_{S})\oplus {\rm ad}(G_{S_i})\oplus (\oplus_j U_j\otimes {(U_i)}_j)
\ea
with $U_j,{(U_i)}_j$ being the irreducible representations of $G_S, G_{S_i}$.  In the simple case of $G_{S}=SU(n)$,
$G_{S_i}=SU(m)$,  and $G_{\Sigma_i}=SU(n+m)$ for example, the chiral ${\cal N}=1$ multiplet is
the bifundamental $(n,\ov{m})$.

We assume that a non-trivial background gauge field configuration acquires a value in a subgroup
$H_S\subset G_S$ and similarly in $H_{S_i}\subset G_{S_i}$. If $G_S\supset\Gamma_S\times H_S$ and
 $G_{S_i}\supset\Gamma_{S_1}\times H_{S_i}$, with $\Gamma_{S,S_i}$ being the corresponding maximal
 $G_{S,S_i}$ subgroups,  the $G_S\times G_{S_i}$ symmetry breaks to the commutant
group $\Gamma =\Gamma_S\times \Gamma_{S_i}$. Denoting also $H=H_S\times H_{S_i}$, the decomposition
of $U\otimes {U_i}$ into irreducible representations of  $\Gamma\times H$ give
\ba
U\otimes {U_i} = \oplus_j (r_j,R_j)
\ea
Let us see how this works for $G_S={\cal E}_6$~\cite{Beasley:2008dc,Callaghan:2011jj}.
Choosing a $H_S=U(1)$  flux, ${\cal E}_6$ breaks to $SO(10)\times U(1)$.
If we set $G_{\Sigma_1}=SU(3)$ and recall the breaking pattern
 $${\cal E}_8\supset{\cal E}_6\times SU(3)\supset SO(10)\times U(1)\times SU(3)$$ we have
 the decomposition of the ${\cal E}_8$ adjoint
 \ba
 248&\ra& (78,1)+(27,3)+(\overline{27},\overline{3})+(1,8)
 \ea
Matter and Higgs fields are  found on matter curves and in the bulk, in the
following representations
\ba
(27,3)&\ra&(1,3)_4+(10,3)_{-2}+(16,3)_1\label{27}\\
(\overline{27},\overline{3})&\ra&(1,\overline{3})_{-4}+(10,\overline{3})_{2}+(\ov{16},\overline{3})_{-1}
\label{27b}\\
( 78,1)&\ra& (45,1)_0+(1,1)_0+(16,1)_{-3}+(\overline{16},1)_3\label{78}
\ea
The net number of chiral fermions  in a specific representation is given as before by $n_{r_j}-n_{r^*_j}$.
In the case of an algebraic curve $\Sigma_i$ the Euler character is written
 as a function of the genus $g$ of the Riemann surface $\Sigma_i$ and the first Chern
 class~\cite{Beasley:2008dc}:
 \ba
 n_{r_j}-n_{r^*_j} &=&(1-g)\,{\rm rk}\,(\Sigma_i,K_{\Sigma_i}^{1/2}\otimes {\cal R}_j)+
 \int_{\Sigma_i}c_1(\Sigma_i,K_{\Sigma_i}^{1/2}\otimes {\cal R}_j)
 \ea
 with $K_{\Sigma_i}^{1/2}$ being the spin bundle over $\Sigma_i$ and ${\cal R}_j$ the
corresponding bundle which transforms as a representation $R_j$.
A recent analysis on ${\cal E}_6$ can be found in~\cite{Callaghan:2011jj}.

\subsection{Two or... three things we should know about  Del Pezzo surfaces}

Since the role of the compact surface $S$ is pivotal for the properties of the model,
let us review a few things about them.

$\bullet$
We are mainly interested  to del Pezzo surfaces.
The simplest ones are $\mathbb{P}^1\times \mathbb{P}^1$ ($= \mathbb{F}_0$ e.g. a Hirzebrough surface~\footnote{
A Hirzebruch surface  is a $\mathbb{P}^1$ fibration over a  $\mathbb{P}^1$;
the general type is classified by an integer index $n$, and denoted by $\mathbb{F}_n$.
It is spanned by two generators $ {\cal S}, {\cal E}$ with the properties
${\cal S}\cdot {\cal S}=-n,\;{\cal S}\cdot {\cal E}=1,\; {\cal E}\cdot {\cal E}=0$.
The canonical divisor (and Chern class) is given by $K_S=-c_1(S)=-2 {\cal S}-(n+2) {\cal E}$
and any effective class is a combination $a {\cal S}+b{\cal E}$, with $a,b\ge 0$.})
and  $dP_0=\mathbb{P}^2$.  The are eight more  del Pezzo surfaces $dP_n$ constructed from an operation known
 as `blow up' of $\mathbb{P}^2$ at generic points. To blow-up a surface (manifold) at a
 marked point, we remove the point and replace it with a line gluing it
in such a way so that we still get a manifold. The points of this line correspond to different
directions from the marked point on the plane.  Del Pezzo surfaces are obtained by applying the
`blow-up' operation up to eight points on the plane.
 A $dP_n$ is generated by the hyperplane divisor $H$ from $\mathbb{P}^2$ and
 the exceptional divisors $E_{1,...,8}$ with intersection numbers
 \ba
 H\cdot H=1,\;H\cdot E_i=0,\;E_i\cdot E_j=-\delta_{ij}\label{IN}
 \ea
The canonical divisor (and the first Chern class $c_1(dP_n)$) is given by
\ba
K_S&=&-c_1(dP_n)=-3H+\sum_{i=1}^nE_i\label{KD}
\ea
The  $dP_n$  generators $C_i$ are given in  Table \ref{dpn1}.

$\bullet$ The effective class $C$ of a curve can be written as
a sum of the generators $C_i$, $C=\sum_i n_iC_i$ for  $n_i>0$.
\begin{table}[!t]
\centering
\renewcommand{\arraystretch}{1.2}
\begin{tabular}{|c|c|c|c|}
\hline
${ S}$& Generators $C_i$&Indices&$\#$ of $C_i$\\
\hline
 $dP_1$   &$ E_1,H-E_1$  &1&2  \\
 \hline
  $dP_2$   &$ E_i, H-E_1-E_2 $&$i=1,2$&3\\
   \hline
  $dP_3$   &$ E_i, H-E_i-E_j$&$i,j=1,2,3$&6\\
    \hline
  $dP_4$   &$ E_i, H-E_i-E_j$&$i,j=1,2,3,4$&10\\
\hline
 $dP_5$   &$ E_i, H-E_i-E_j,2H-E_i-E_j-E_k-E_l-E_m$&$i,j,\dots=1,\dots,5$&16\\
\hline
  $\cdots$   &$ \cdots$&$\cdots$&$\cdots$\\
\hline
 $dP_8$   & $E_k,\;H-E_{k}-E_l,\; 2H-\sum_{j=1}^5E_{n_j}$&$k,l,...=1,\dots,8$&240\\
          & $ 3H-2 E_k-\sum_{j=1}^6E_{n_j},\; 4H-2 (E_k+E_l+E_m)-\sum_{j=1}^5E_{n_j}$&&\\
 \hline
\end{tabular}
 \caption{The generators  of a few del Pezzo surfaces (see~\cite{Heckman:2009mn}). All effective classes can
 be written as linear combinations of $C_i$ with coefficients non-negative integers.\label{dpn1} }
\end{table}
The characteristic property of a del Pezzo surface is that $c_1$ is ample, that is, it has positive
intersection with every effective curve.
This in particular implies that $K$ must have
positive self-intersection,
\[ K_S\cdot K_S=9-n\]
which gives the restriction $n\le 8$.

A K\"ahler class can be defined as follows
\[\omega=A\,H-\sum_{i=1}^na_iE_i\]
For a line bundle $L$ on del Pezzo with $c_1(L)=\sum_{i=1}^nm_iE_i$, ($m_i$ integers)
the condition $\omega\cdot c_1(L)=0$  implies
\[ \sum_ia_im_i=0\]
while for sufficiently large $A$, for any divisor $D$, the intersection is
positive $\omega\cdot D>0$.

$\bullet$ To see the connection of $dP_n$ with exceptional algebras
 let's define the generators (for $n\ge 3$)
\ba
a_1=E_1-E_2,\,\dots, \, a_{n-1}=E_{n-1}-E_n,\, a_n=H-E_1-E_2-E_3
\ea
Using the dot product for the  $E_i,H$ generators, we get
\ba
 a_i\cdot a_j=2\delta_{ij}-\delta_{i,j+1}-\delta_{j,i+1}=
\left\{\begin{array}{rl}2&i=j\\-1&i=j+1\\-1&j=i+1\end{array}\right.
\label{Eroot}
\ea
The intersection product of $a_i$'s is identical to minus the Cartan matrix for the dot
product of the simple roots of the corresponding algebra $E_n$. In the particular case
of $dP_2$ there is only one generator $E_1-E_2$ which is identified as a root of $SU(2)$.

\subsection{The $SU(5)$ model}
The F-theory derivation of the $SU(5)$  model has attracted the interest of many authors.
So, let us consider now  that we have a singularity  $G_S=SU(5)$. We assume that
the gauge symmetry  breaks  to SM by turning on a $U(1)_Y$ flux.
We write the decomposition of the $SU(5)$ gauge multiplet as
\[24\ra R_0+R_{-5/6}+R_{5/6}\]
where
\ba
R_0=(8,1)_0+(1,3)_0+(1,1)_0,\;R_{-5/6}=(3,2)_{-5/6},\;R_{5/6}=(\bar 3,2)_{5/6} \cdot
\label{24Ri}
\ea
As we have seen above, massless fields  in the bulk are given by  the Euler characteristic $\chi$.
 In order to avoid the massless exotics  $R_{\pm 5/6}$ we impose the condition
 ${\chi}(S,L^{\pm 5/6})=0$.  Taking the difference
 \[0={\chi}(S,L^{ 5/6})-{\chi}(S,L^{- 5/6})=c_1(L^{5/6})\cdot c_1(dP_8)=-c_1(L^{5/6})\cdot K_S\]
so $c_1(L^{5/6})\cdot K_S=0$ which means that $c_1(L^{5/6})$ is orthogonal to $K_S$, i.e.  it
is  a vector in the orthogonal complement of the canonical class.
Substituting to the Euler character we find
\ba
\chi(S,L^{5/6})=0\;&\Rightarrow&\;1+\frac 12 c_1(L^{5/6})\cdot c_1(L^{5/6})=0
\label{noexotic}
\ea
The vanishing of the latter implies
\ba
c_1(L^{5/6})\cdot c_1(L^{5/6})&=&-2
\ea
This is the condition for $c_1(L^{5/6})$ to correspond to a root of $E_n$ (see $(\ref{Eroot})$)
while it implies a fractional line bundle $L={\cal O}(E_i-E_j)^{1/5}$ (yet consistent with
bulk gauge field configurations~\cite{Beasley:2008kw}).

To obtain the chiral and Higgs spectrum, we should consider the intersections with other branes.
Recall that chiral matter and Higgs fields reside on the $10$ and $5,\bar 5$ representations.

\begin{figure}[tbh]
\centering
\includegraphics[scale=0.68]{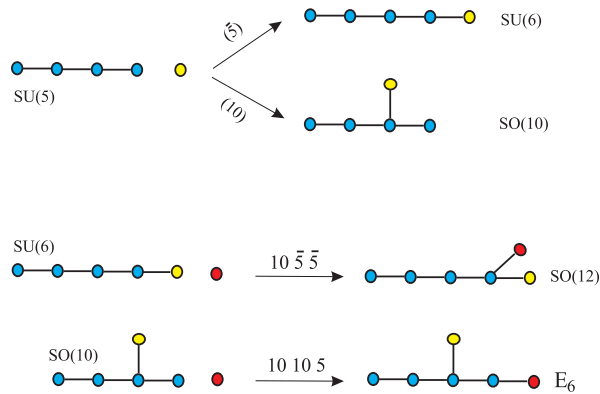}
\caption{Enhancements of the $A_4$ singularity at double and triple intersections.}
 \label{Dynk}
\end{figure}

In the F-theory set up,  the $5$ and $\bar 5$ reside on curves where  $SU(5)$ enhances to $SU(6)$.
Similarly the $10$'s are localised on curves where $SU(5)$ enhances to $SO(10)$. When three of these matter
 curves meet at one point, a trilinear Yukawa coupling is generated while the gauge symmetry is further
 enhanced. There is a pellucid  way to see these enhancements with the help of Dynkin diagrams.
In figure~\ref{Dynk}  we start with an $A_4$ singularity which corresponds to $SU(5)$.
There are two ways to extend this diagram:
in the first one we observe that the symmetry is enhanced to $SU(6)$ and the
$5$-representation of $SU(5)$ is found in the decomposition of the $SU(6)$ adjoint,
${35}\to  24_0+1_0+{5_6}+{\bar 5_{-6}}$. In the second case we observe from figure~\ref{Dynk}
that we can also have an $SO(10)$ enhancement while the $10$ of $SU(5)$ is found
in the adjoint decomposition  $45\to 24_0+1_0+{10_4}+{\overline{10}_{-4}}$.
 The top Yukawa coupling $10\cdot 10\cdot 5$ originates from an
 ${\cal E}_6$ enhancement and the bottom $10\cdot\bar 5 \cdot\bar  5$ from an $SO(12)$.

For an implementation of the above, consider the particular case of $SU(5)$ toy model discussed in
 section 17  of ref~\cite{Beasley:2008kw}. The  surface $S$ is of the del Pezzo type   $dP_8$
 which is generated by the hyperplane divisor $H$ from
$\mathbb{P}^2$ and the exceptional divisors $E_{1,...,8}$ with intersection numbers
and canonical divisor  for $dP_8$ given by (\ref{IN}) and (\ref{KD}).
Denoting with $C$ and $g$ the class and the genus of a given matter  curve respectively, one has
\ba
C\cdot (C+K_S)&=&2 g-2\label{genus}
\ea
In this particular example the $10_M$  chiral matter of the three generations resides on one
$\Sigma_{10}$, with $C=2 H-E_1-E_5$
and the three $\bar 5_M$  on a single  $\Sigma_{5}^1$ curve with $C=H$. Higgs fields
$5_H$ and $\bar 5_{\bar H}$ reside on different  $\Sigma_{5}^{2,3}$ matter curves
with classes $C= H-E_1-E_3$ and $ H-E_2-E_4$ respectively, giving $g=0$ for all curves, three
families and a Higgs pair.
Further details of this model can be worked out using the properties of $dP_8$ and can
be found in~\cite{Beasley:2008kw}.

Next we will discuss in detail the $SU(5)$ model and other cases in the spectral cover picture.

\section{Spectral cover approach}

An equivalent  description of the supersymmetric configurations of the 8-dimensional gauge theory
can be given in terms of adjoint scalars and gauge fields, corresponding  to the so called
Higgs bundle picture \cite{Donagi:2008ca}. In the spectral cover picture we  concentrate
in the vicinity  of the  chosen surface $S$ associated to the GUT group $G_S$,
while its neighborhood is described by a spectral surface. The intersections of
the spectral cover with the surface $S$ encode the information  about the
spectrum and its properties.

In local F-theory models we consider  the maximum singularity of the elliptic fibrations,
i.e. ${\cal E}_8$,  thus assuming that our effective theory has a GUT group $G_S$ the spectral
 cover group corresponds to its commutant with respect to ${\cal E}_8$.
We recall that all viable  gauge groups $G_S$ embedded in ${\cal E}_8$,
can be inferred by the embedding formula~\cite{Beasley:2008dc}
\ba
\frac{{\cal E}_n\times SU(m)}{Z_m}\subset {\cal E}_8,\;\; n+m=9
\ea
 Of particular interest are the cases where $G_S$ is one of the
 phenomenologically viable GUTs ${\cal E}_6, SO(10)$ or $SU(5)$. The corresponding
 decompositions are
\ba
{\cal E}_8&\supset&{\cal E}_6\times SU(3)\ra {\cal E}_6\times U(1)^3\ra [SO(10)\times U(1)]\times U(1)^2\label{su3}\\
{\cal E}_8&\supset&{\cal E}_5\{=SO(10)\}\times SU(4)\label{E5chain}\\
    &\ra& [SU(5)\times U(1)]\times SU(4)\ra [SU(5)\times U(1)]\times U(1)^3\label{su4}\nn\\
{\cal E}_8&\supset&SU(5)\times SU(5)_{\perp}\ra SU(5)\times U(1)^4\label{su5}
\ea
A complete list of all possibilities can be found in~\cite{Beasley:2008dc}.
Here we will construct the $SU(5)$ and flipped $SU(5)$ models, while similar analysis
for the ${\cal E}_6$ model can be found in~\cite{Callaghan:2011jj}.

\subsection{ The $SU(5)$ model from the spectral cover}

If we take $G_S=SU(5)$ the corresponding spectral surface is  $SU(5)_{\perp}$.
Matter resides in  the adjoint representation of ${\cal E}_8$ which in this case
 decomposes as
\[248 = (24, 1) + (1, 24) + (5, 10) + (\bar 5, \overline{10}) + (10, \bar 5) + (\overline{10}, 5)\]
The decomposition appears under $SU(5)_{GUT}\times SU(5)_{\perp}$ where the $
SU(5)_{\perp}$ is the  group describing the  bundle in the vicinity.

We label the weights of  $SU(5)_{\perp}$ with $t_i$ subject to $ \sum_{i=1}^5t_i=0$,
while we assume further breaking of $SU(5)_{\perp}$ to
\[SU(5)_{\perp}\ra U(1)^4\]
Thus the $10$ representations of $SU(5)_{GUT}$ originate from the $(10, \bar 5)$ component
they reside on matter curves $\Sigma_{10_{t_i}}$
and are characterised by  the  weights ${t_i}$. Similarly, the $5/\bar 5$ representations
reside on $\Sigma_{5_{t_i+t_j}}$.

The corresponding spectral cover equation is obtained by defining the  homogeneous coordinates
\[z\ra U,\; x\ra V^2,\; y\ra V^3\]
 so that the Weierstrass equation becomes
\[ 0=b_0 { U^5}+b_2 V^2 { U^3}+b_3V^3U^2+b_4V^4 {U}+b_5V^5\]
with $U,V$ being sections of $-t$ and $c_1-t$ respectively. We can turn  this equation to a
fifth degree polynomial in terms of  the affine parameter
$s=U/V$:
\[P_5=\sum_{k=0}^5 b_ks^{5-k}\;=\;b_5+b_4s+b_3s^2+b_2s^3+b_1s^4+b_0s^5\]
where we have divided by the fifth power $V^5$, so that each term in the
last equation becomes section of $c_1-t$.
The roots of the spectral cover equation~\cite{Donagi:2008ca,Marsano:2008jq}
\ba
0&=&b_5+b_4s+b_3s^2+b_2s^3+b_0s^5\propto \prod_{i=1}^5(s+t_i)\label{spec5}
\ea
are identified as the $SU(5)$ weights  $t_i$.

In the above the coefficient $b_1$ is taken to be zero since it corresponds to the sum
of the roots which for $SU(n)$ is always zero, $\sum t_i=0$.
Also, it can be seen that the coefficient $b_5$ is equal to the product of
the roots, i.e. $b_5=t_1t_2t_3t_4t_5$ and the  $\Sigma_{10}$ curves where
 the corresponding matter multiplets are  localised are determined by
 the five zeros
\ba
\Sigma_{10_i},\;\;\; b_5=\prod_{=i=1}^5t_i=0\ra t_i=0,\;\;\; i=1,2,3,4,5\label{tens}
\ea

The model effectively appears with a symmetry $SU(5)_{GUT}\times U(1)^4$. In order to write
a Yukawa term, this symmetry should be respected.  Thus, writing the coupling involving the up quark masses
\[ {\cal W}\supset 10_{t_i}\,10_{t_j}\,5_{-t_i-t_j}\]
 would  appear to involve two different generations. On the other hand, phenomenology requires a rank one mass
 matrix at tree-level to account for the heavy top mass.  A similar conclusion holds for the bottom
 mass term.  More generally,   the known hierarchical fermion mass spectrum and the heaviness of the
  third generation however, is compatible with rank one structure of the mass matrices at tree-level.
 This requires a solution where at least two of the curves are identified through some (discrete) symmetry.

 This idea of identification is corroborated also by the following fact.
In the spectral cover approach, we have seen that the properties of
the manifold are encoded into the coefficients $b_i$. Matter curves
on the other hand are associated to the roots $t_i$ which are
polynomial solutions with factors combinations of  $b_i$'s, thus
\[ b_i=b_i(t_j)\]
Generically, the inversion of these equations will lead to branchcuts.
The solutions $t_j=t_j(b_i)$ are then subject to monodromy actions.

To get a feeling of the procedure we present an example (given in~\cite{Heckman:2009mn})
by considering the simplest case of the ${\cal Z}_2$ monodromy. Suppose that two of the
roots in (\ref{spec5}) do not factorize. This implies that the second degree polynomial
\[a_1+a_2s+a_3s^2=0\]
 cannot be expressed in simple polynomials of the base coordinates.
 The solutions can be written
\[ s_1=\frac{-a_2+\sqrt{w}}{2a_3},\; s_2=\frac{-a_2-\sqrt{w}}{2a_3}\]
with $w=a_2^2-4a_1a_3$.  These exhibit branchcuts and since
\[\sqrt{w}=e^{i\theta/2}\sqrt{|w|}\]
under a $2\pi$ rotation around the brane configuration $\theta\ra \theta+2\pi$
we get $\sqrt{w}\ra -\sqrt{w}$ and
\[s_1\leftrightarrow s_2\]
This means that the two branes interchange locations $s=s_1$ and $s=s_2$. This is
equivalent of taking the quotient of the parent theory with a ${\cal Z}_2$ symmetry.
If this is among $t_1\leftrightarrow t_2$ the coupling now reads
\[ {\cal W}\supset 10_{t_1}\,10_{t_2}\,5_{-t_1-t_2}\ra 10_{t_1}\,10_{t_1}\,5_{-2t_1}\]
providing a diagonal mass term since the two curves are  identified.

Since the $SU(5)$ spectral cover is described by the $5$-degree polynomial shown above,
 the various monodromy actions are associated to the  possible ways of splitting
 the polynomial according to
\ba
{\cal Z}_2:\;&&({a_1+a_2s+a_3s^2})(a_4+a_5s)(a_6+a_7s)(a_8+a_9s)\nn\\
{\cal Z}_2\times{\cal Z}_2: &&({ a_1+a_2s+a_3s^2})({a_4+a_5s+a_6s^2})(a_7+a_8s)\nn\\
{\cal Z}_3:&&({ a_1+a_2s+a_3s^2+a_4s^3})(a_5+a_6s)(a_7+a_8s)\nn\\
{\cal Z}_4:&&({ a_1+a_2s+a_3s^2+a_4s^3+a_5s^4})(a_6+a_7s)\nn\\
{\cal Z}_3\times{\cal Z}_2:&&({ a_1+a_2s+a_3s^2+a_4s^3})(a_5+a_6s+a_7s^2)\nn\\
{\rm no\, split}:&&({ a_1+a_2s+a_3s^2+a_4s^4+a_5s^5})\nn
\ea

\section{The case of ${\cal Z}_2$ monodromy}

Up to this point we have discussed  the constraints from the gauge symmetry $G_S$ that should
be imposed on the Yukawa sector of the effective field theory.  We have seen that
the $U(1)$ factors are not entirely independent since they undergo a series
of monodromies. In general, the theory must be the quotient by some monodromy
group which leaves the roots of the gauge symmetry $G_S$ invariant.
 In the following we attempt to implement the constraints
obtained from the previous symmetry breaking stages into the $SU(5)_{GUT}$
model imposing a ${\cal Z}_2$ monodromy among  $t_1, t_2$.
Expanding, we may determine the homology class for each of the coefficients $a_i$ by comparison with
the $b_k$'s. Thus, one gets
\begin{equation}
\label{bfroma}
\begin{split}
b_0&=a_3 a_5 a_7 a_9
\\
b_1&= a_3 a_5 a_7 a_8+a_3 a_4 a_9 a_7+a_2 a_5 a_7 a_9 +a_3 a_5 a_6 a_9\\\
b_2&=a_3 a_5 a_6 a_8+a_2 a_5 a_8 a_7+a_2 a_5 a_9 a_6+a_1 a_5 a_9 a_7+a_3 a_4 a_7 a_8+a_3 a_4 a_6
   a_9+a_2 a_4 a_7 a_9\\
   b_3&=a_3 a_4 a_8 a_6+a_2 a_5 a_8 a_6+a_2 a_4 a_8 a_7+a_1 a_7 a_8 a_5+a_2 a_4 a_6 a_9+a_1 a_5 a_6  a_9+a_1 a_4 a_7 a_9\\
   b_4&=a_2 a_4 a_8 a_6+a_1 a_5 a_8 a_6+a_1 a_4 a_8 a_7+a_1 a_4 a_6 a_9\\
   b_5&=a_1a_4a_6a_8
   \end{split}
   \end{equation}
We first solve the constraint $b_1=0$.  We make the Ansatz~\cite{Dudas:2010zb}
\[ a_2 =-c ( a_5 a_7 a_8+a_4 a_9 a_7+ a_5 a_6 a_9),\; a_3=c  a_5 a_7 a_9\]
Substituting into $b_n$'s we get
\ba
b_0&=&c\, a_5^2 a_7^2 a_9^2\nn \\
b_2&=&a_1 a_5 a_7 a_9-\left(a_5^2 a_7^2 a_8^2+a_5 a_7 \left(a_5 a_6+a_4 a_7\right) a_9 a_8+\left(a_5^2 a_6^2+a_4 a_5 a_7 a_6+a_4^2 a_7^2\right) a_9^2\right) c
\nn\\
b_3&=&a_1 \left(a_5 a_7 a_8+a_5 a_6 a_9+a_4 a_7 a_9\right)-\left(a_5 a_6+a_4 a_7\right) \left(a_5 a_8+a_4 a_9\right) \left(a_7 a_8+a_6 a_9\right) c\nn\\
b_4&=&a_1 \left(a_5 a_6 a_8+a_4 a_7 a_8+a_4 a_6 a_9\right)-a_4 a_6 a_8 \left(a_5 a_7 a_8+a_5 a_6 a_9+a_4 a_7 a_9\right) c\nn\\
b_5&=&a_1 a_4 a_6 a_8\nn
\ea
Next, we observe that we have  to determine the homology classes $[a_i]$ of nine unknowns $a_1,\dots a_9$
   in terms of the $b_k$-classes $[b_k]$.  From  (\ref{bfroma}) we deduce that the latter satisfy the general
   equation $[b_k]=[a_l]+[a_m]+[a_n]+[a_p]$ for $k+l+m+n+p=24$.   Three classes are left  unspecified which we
   choose them to be $[a_l]=\chi_l,l=5,7,9$. The rest are computed easily and presented in Table
   \ref{a_class}.

The $\Sigma_{10}$ curves are found setting $s=0$ in the polynomial
\ba b_5\equiv\Pi_5(0)=a_1a_4a_5a_6=0&\ra& a_1=0,\, a_4=0,\, a_5=0, \,a_6=0\ea
Thus, after the monodromy action, we obtain four curves (one less compared to no-monodromy case)
to arrange the appropriate pieces of the three (3) families.
\begin{table}
\begin{center}
\begin{tabular}{c|c|c|c|c|c|c|c|c}
\hline
$a_1$& $a_2$& $a_3$& $a_4$& $a_5$& $a_6$& $a_7$&$a_8$&$a_9$
\\
$\eta-2c_1-\chi$& $\eta-c_1-\chi$& $\eta-\chi$& $-c_1+x_5$& $x_5$& $-c_1+x_7$& $x_7$&$-c_1+\chi_9$&$\chi_9$\\
\hline
\end{tabular}
\end{center}
\caption{Homology classes for coefficients $a_i$  for the ${\cal Z}_2$ ($SU(5)$) case}
\label{a_class}
\end{table}
The $\Sigma_{5}$ curves are treated similarly.
To determine the properties of the fiveplets we need the corresponding spectral cover equation.
This is a 10-degree polynomial
\[ {\cal P}_{10}(s)=\sum_{n=1}^{10}c_ns^{10-n}=b_0\prod_{i,j}(s-t_i-t_j),\; i<j,\; i,j=1,\dots, 5\]
We can convert the coefficients $c_n=c_n(t_j)$  to functions of
$c_n(b_j)$.  In particular we are interested for the  value ${\cal P}_{10}(0)$
given by the coefficient $c_{10}$ which can be expressed in terms of $b_k$ according to
\[c_{10}(b_k)= b_3^2b_4-b_2b_3b_5+b_0b_5^2=0\]
Using the equations $b_k(a_i)$ and the Ansatz, we can split this equation into seven factors
which correspond to the seven distinct fiveplets left after the ${\cal Z}_2$ monodromy action.
\be
\begin{split}
P_5&= \left(a_1-c a_4 \left(a_7 a_8+a_6 a_9\right)\right)\times \left(a_1-c \left(a_5 a_6+a_4 a_7\right) a_8\right)\times \left(a_1-c a_6 \left(a_5 a_8+a_4 a_9\right)\right)\\
   &\times  \left(a_4 a_7 a_9+a_5 \left(a_7 a_8+a_6
   a_9\right)\right)\times \left(a_5 a_6+a_4 a_7\right)\times \left(a_5 a_8+a_4
   a_9\right) \left(a_7 a_8+a_6 a_9\right)
   \end{split}
   \ee
Their homologies can be specified using those of $a_i$. Notice that in the first line of the above
the three factors correspond to three fiveplets of the same homology class $[a_1]=\eta-2c_1-\chi$.
The complete spectrum is presented in Table~\ref{Reps}.
\begin{table}[tbp] \centering%
\begin{tabular}{|l|c|c|c|c|}
\hline
Field&$U(1)_i$& homology& $U(1)_Y$-flux&$U(1)$-flux\\
\hline
$10^{(1)}=10_3$& $t_{1,2}$& $\eta-2c_1-{\chi}$&$ -N$ &$M_{10_1}$\\ \hline
$10^{(2)}=10_1$& $t_{3}$& $-c_1+\chi_7$&$ N_7$ &$M_{10_2}$\\ \hline
$10^{(3)}=10_2$& $t_{4}$& $-c_1+\chi_8$&$ N_8$ &$M_{10_3}$\\ \hline
$10^{(4)}=10_2'$& $t_{5}$& $-c_1+\chi_9$&$ N_9$ &$M_{10_4}$\\ \hline
$5^{(0)}=5_{h_u}$& $-t_{1}-t_2$& $-c_1+{\chi}$&$ N$ &$M_{5_{h_u}}$\\ \hline
$5^{(1)}=5_2$& $-t_{1,2}-t_3$& $\eta -2c_1-{\chi}$&$ -N$ &$M_{5_1}$\\ \hline
$5^{(2)}=5_3$& $-t_{1,2}-t_4$& $\eta -2c_1-{\chi}$&$ -N$ &$M_{5_2}$\\ \hline
$5^{(3)}=5_x$& $-t_{1,2}-t_5$& $\eta -2c_1-{\chi}$&$ -N$ &$M_{5_3}$\\ \hline
$5^{(4)}=5_1$& $-t_{3}-t_4$& $-c_1+{\chi}-\chi_9$&$N-N_9$ &$M_{5_4}$\\ \hline
$5^{(5)}=5_{h_d}$& $-t_{3}-t_5$& $-c_1+{\chi}-\chi_8$&$ N-N_8$ &$M_{5_{h_d}}$\\ \hline
$5^{(6)}=5_y$& $-t_{4}-t_5$& $-c_1+{\chi}-\chi_7$&$ N-N_7$ &$M_{5_6}$\\ \hline
\end{tabular}%
\caption{Field representation content under $SU(5)\times U(1)_{t_i}$, their homology
class and flux restrictions~\cite{Dudas:2010zb} for the model~\cite{Leontaris:2010zd}.
Superscripts in the first column are numbering the curves, while subscripts indicate the
family, the Higgs etc. For convenience, only the properties of $10,5$ are shown.
$\ov{10},\ov{5}$ are characterized by opposite values of $t_i\ra -t_i$ etc.
Note that the fluxes satisfy $N=N_7 +  N_8 + N_9$ and $\sum_iM_{10_i}+\sum_jM_{5_j}=0$
while  ${\chi}=\chi_7 +  \chi_8 + \chi_9$.}
\label{Reps}
\end{table}
Recall now that the $SU(5)$ multiplets decompose to Standard Model multiplets according to
\begin{equation}
\label{105split}
\begin{split}
10&\ra (3,2)_{\frac 16}+(\bar 3,1)_{-\frac 23}+(1,1)_1\to (Q,u^c,e^c)\\
5&\ra (3,1)_{-\frac 13}+(1,2)_{\frac 12}\to (d^c,\ell)
\end{split}
\end{equation}
We have pointed out that
in F-theory constructions one of the possible ways to break the GUT symmetry is to turn on a flux on the
worldvolume of the seven-brane supporting the unified gauge group.  In the present case, the $SU(5)$ gauge
symmetry can be broken by turning on a non-trivial flux along the hypercharge
with $Q_Y={\rm diag}\{-\frac 13,-\frac 13,-\frac 13,\frac 12,\frac 12\}$.  As a result,  $SU(5)$
multiplets residing on certain curves where the flux restricts non-trivially, might split.
This means that some SM pieces of the (\ref{105split}) decomposition could be swept away by flux.
 In the case of
Higgs fiveplets in particular, this mechanism could be used to remove the unwanted triplets.
To implement this idea in a specific scenario, we recall  first the  $SU(5)$ embedding to ${\cal E}_8$
\[
{\cal E}_8\ra SU(5)_{GUT}\times U(1)^4
\]
The $SU(5)$ chiral and Higgs matter fields descend from the adjoint representation of the ${\cal E}_8$ symmetry
and reside on the various curves denoted with $\Sigma_{10_j},\Sigma_{\bar 5_i}$.
Suppose that $M_{10_j},M_{5_i}$  are two integers representing the number of  $10$ and $\bar 5$ representations
in a specific construction. The $U(1)$ fluxes (those not included in $SU(5)_{GUT}$) together with the
tracelessness condition $\sum_iF_{U(1)_i}=0$ imply the  following condition on the numbers of multiplets~\cite{Dudas:2010zb,Marsano:2010sq}
\begin{equation}
\label{trace0}
\sum_i M_5^i+\sum_j M_{10}^j=0
\end{equation}
Consider first the case that we have all $10$-type chiral matter  accommodated only on one $\Sigma_{10}$ curve
and all chiral states $\bar{5}$ respectively
on a single $\Sigma_{\bar 5}$ curve. Then condition (\ref{trace0}) implies the relation $M_{10}=-M_5=M$.

We denote with $N_{Y_5}, N_{Y_{10}}$ the corresponding units of $Y$ flux which splits the
$SU(5)$ multiplets according to
\begin{equation}
\label{model_dif}
\begin{split}
\Sigma_{\bar 5}:\left\{
\begin{array}{l}
n_{(\mathbf{3,1})_{-1/3}}-n_{(\mathbf{\bar{3},1})_{1/3}}=M_5\\
n_{(\mathbf{1,2})_{1/2}}-n_{(\mathbf{1,2})_{-1/2}}=M_5+N_{Y_5}
\end{array}
\right.
\Sigma_{10}:\left\{
\begin{array}{l}
n_{(\mathbf{3,2})_{1/6}}-n_{(\mathbf{\bar{3},2})_{-1/6}}=M_{10}\\
n_{(\mathbf{\bar{3},1})_{-2/3}}-n_{(\mathbf{3,1})_{2/3}}=M_{10}-N_{Y_{10}}\\
n_{(\mathbf{1,1})_{1}}-n_{(\mathbf{1,1})_{-1}}\hspace*{.53cm}=M_{10}+N_{Y_{10}}
\end{array}
\right.
\end{split}
\end{equation}
Notice that these formulae count the number of $5$-components minus those of $\bar 5$
and the number of $10$ components minus those of $\overline{10}$. Since we know that
families are accommodated on $\bar 5$'s
 we require  $n_{(\mathbf{\bar{3},1})_{1/3}}>n_{(\mathbf{3,1})_{-1/3}}$ which implies
$M_5<0$. Similarly, because the remaining pieces of fermion generations live on $10$'s,
 we wish to end up with $10$-components after
the symmetry breaking, hence we should have $M_{10}>0$. For example, for exactly three generations
we should demand $M_{10}=-M_{5}=3$ and $N_{Y_j}=0$.  In general  various curves  belong to
different homology classes and flux restricts non-trivially to some of them, thus $N_{Y_j}\ne 0$
at least for some values of $j$.

\subsection{A realistic model with Doublet-Triplet splitting}

We will discuss here  the model of~\cite{Leontaris:2010zd} which emerges from the general class~\cite{Dudas:2010zb}
presented in  Table \ref{Reps}.  The first two columns give the field content under $SU(5)\times U(1)_{t_i}$ for the
case of ${\cal Z}_{2}$ monodromy.  The third column presents the homology classes expressed in terms of  $c_{1},\eta$ and the $\chi_{i}$ the latter being  unspecified subject only to the condition ${\chi}=\chi_7 +  \chi_8 + \chi_9$.  If  ${\cal F}_{Y}$ denotes the $U(1)_Y$ flux,  to avoid a Green-Schwarz mass for the corresponding gauge boson we must require ${\cal F}_{Y}\cdot\eta={\cal F}_{Y}\cdot c_1=0$. Then,  we get $N_i={\cal F}_{Y}\cdot\chi_i$ and  consequently $N={\cal F}_{Y}\cdot \chi= N_7+N_8+N_9$.  Using these facts, all remaining entries of column 4 in Table \ref{Reps} are easily deduced.

We now take the flux parameters to be  $M_{10_{1,2,3}}=1$, $M_{5_{1,2,4}}=-1$
 and $N=0$, while we have the freedom to choose $N_{7,8,9}$ subject only to the constraint $N=N_7+N_8+N_9$.
This choice of $M_i,N_j$'s ensures the existence of three $10$ and three $\bar 5$ representations which are needed to
accommodate the three chiral families.

Next we use the $U(1)_Y$ flux mechanism to realise the doublet triplet splitting and
 make the model free from dangerous color triplets at scales below $M_{GUT}$. We choose
 $M_{5_{h_u}}=1$, to accommodate the Higgs  ${5}_{h_{u}}$. In addition we choose
  $M_{5_{h_d}}=0$ and $N_8=1$  so that we are left only with the $h_d$-doublet in the
  corresponding Higgs fiveplet
\ba
\Sigma_{5_{h_d}}:\left\{\begin{array}{ll} n_{(3,1)_{ - 1/3} }  -
 n_{(\overline 3 ,1)_{1/3} }& = M_{5_5 }  = 0
\\
n_{(1,2)_{1/2} }  - n_{(1,2)_{ - 1/2} } & = M_{5_5 }  + N - N_8  =  - 1
\end{array}\right.
\ea
In order to  satisfy the trace conditions we  choose $M_{5_6 }  =  - 1,
\;N_7  =  - 1$ so that $\bar{5}^{(6)}$ has only a colour triplet component:
\ba
\Sigma_{5^{(6)}}:\left\{\begin{array}{ll}n_{(3,1)_{ - 1/3} }  -
n_{(\overline 3 ,1)_{1/3} } & = M_{5_6 }  =  - 1
\\
n_{(1,2)_{1/2} }  - n_{(1,2)_{ - 1/2} } & = M_{5_6 }  + N - N_7  =  0\end{array}\right.
\ea
We observe that in this simple example we have succeeded to disentangle the colour triplet from the Higgs
curve at the price of generating however a new one in a different matter curve. Yet, this
allows the possibility of realising the doublet-triplet splitting since we can
generate a heavy mass $M_{D}$ for the triplet  by coupling it to an  antitriplet via the appropriate
superpotential term~\cite{Leontaris:2010zd}. This way we obtain the corresponding Higgs doublets light.

However from Table \ref{Reps} one may see that the matter on the $\Sigma_{10^{(2,3)}}$ curves will be
affected by the $N_{7,8}$ flux. In particular the content of $10/{\ov 10}$-representations on
$\Sigma_{10^{(2,3)}}$  splits  as follows
\be
\Sigma_{10^{(2)}}:\left\{\begin{array}{ll}
n_{(3,2)_{1/6} }  - n_{(\overline 3 ,2)_{ - 1/6} }  &= M_{10_2 }  = 1
\\
n_{(\overline 3 ,1)_{ - 2/3} }  - n_{(3,1)_{2/3} }  &= M_{10_2 }  - N_7  = 2
\\
n_{(1,1)_1 }  - n_{(1,1)_{ - 1} }  &= M_{10_2 }  + N_7  = 0
\end{array}\right.\ee
\be
\Sigma_{10^{(3)}}:\left\{\begin{array}{ll}
n_{(3,2)_{1/6} }  - n_{(\overline 3 ,2)_{ - 1/6} } & = M_{10_3 }  = 1
\\
n_{(\overline 3 ,1)_{ - 2/3} }  - n_{(3,1)_{2/3} } & = M_{10_3 }  - N_8  = 0
\\
n_{(1,1)_1 }  - n_{(1,1)_{ - 1} } & = M_{10_3 }  + N_8  = 2.
\end{array}\right.
\ee
We observe that in the presence of flux one $e^c=(1,1)_1$ state is `displaced'  from $\Sigma_{10^{(2)}}$ to the
$\Sigma_{10^{(3)}}$ curve.  A similar dislocation  occurs for one $u^c=(\bar 3,1)_{-2/3}$ of $\Sigma_{10^{(3)}}$  which
`reappears' in $\Sigma_{10^{(2)}}$.  We note that this fact implies a different  texture for the up, down and charged lepton mass matrices. It can be checked that the particular distribution of the chiral matter on the
specific  matter curves can lead to interesting results with respect to the
fermion mass structure and other phenomenological properties of the model~\cite{Leontaris:2010zd}.
For clarity, the final distribution of the MSSM spectrum along the available matter curves is summarized  in Table~\ref{content}.
\begin{table}[!t]
\centering
\begin{tabular}{|l|r|r|rrr||l|r|r|rr|}
\hline
\multicolumn{11}{|c|}{Chiral Matter}\\
\hline
\hline
     &  $M$  &  $N$  &  $Q$  &  $u^c$  &  $e^c$ &     &  $M$  &  $N$  &  $d^c$&  $L$    \\
\hline
$10^{(1)}\,(F_3)$ &  1  &  0  &  1  &  1  &  1        &$5^{(4)}\,({\bar f}_1)$  & -1  &  0  &  -1 & -1 \\
$10^{(2)}\,(F_{2,1})$ &  1  &  -1 &  1  &  2  &  0  &$5^{(1)}\,({\bar f}_2)$  &  -1 &  0  & -1  & -1   \\
$10^{(3)}\,(F_{1,2})$ &  1  &  1  &  1  &  0  &  2  &$5^{(2)}\,({\bar f}_3)$  &  -1 &  0  & -1  & -1 \\
$10^{(4)}\,(-)$ &  0  &  0  &  0  &  0  &  0         &$5^{(3)}\,(-)$  & 0   &  0  &  0  &  0   \\
\hline
\end{tabular}

\begin{tabular}{lrrrr}
     &  &    &  &
\end{tabular}

\begin{tabular}{|l|r|r|rr|}
\hline
\multicolumn{5}{|c|}{Higgs and Colour Triplets}\\
\hline
\hline
     &  $M$  &  $N$  &  $T$&  $h_{u,d}$    \\
\hline
$5^{(0)}\,(h_u,T)$&  1    &  0    &  1    &  1\\
\hline
$5^{(5)}\,(h_d)$  & 0   &  -1 &  0  & -1  \\
\hline
$5^{(6)}\,(\bar T)$  & -1  &  1  & -1  & 0\\
\hline
\end{tabular}
\caption{The distribution of the chiral and Higgs matter content of the minimal
model along the available curves, after the $U(1)_Y$ flux is turned on. The three families
  $F_{i}=10_i,\bar{f}_j=\bar 5_j$ are assigned on the curves as  indicated.
The Higgs doublets $h_{u,d}$ and   $T/\bar T$  triplets descend from three different curves.}
\label{content}
\end{table}

We close the section with a few remarks about the $SU(5)$ singlets. These  are found on curves extending
away from the GUT surface $S$. In particular, singlet fields  inhabit on curves in $B_3$ that project down to the curves on the GUT surface~\cite{Marsano:2009wr}.  However, some of  their  properties could in principle be captured by
the corresponding defining equation. Thus, if we work in analogy with the non-abelian representations,
 we could determine their homologies by examining the polynomial equation
 $\prod_{i\ne j} (t_i-t_j)$ in terms of $b_n$'s. The zeroth order term of the polynomial gives~\cite{Callaghan:2011jj}
\ba
P_0&=&3125 b_5^4 b_0^5+256 b_4^5 b_0^4-3750 b_2 b_3 b_5^3 b_0^4+2000 b_2 b_4^2 b_5^2 b_0^4+2250 b_3^2 b_4 b_5^2 b_0^4\nn\\
                &&-1600 b_3 b_4^3 b_5 b_0^4-128 b_2^2 b_4^4 b_0^3+144 b_2 b_3^2 b_4^3 b_0^3-27 b_3^4 b_4^2 b_0^3+825 b_2^2 b_3^2 b_5^2 b_0^3\nn\\
                &&-900 b_2^3 b_4 b_5^2 b_0^3+108 b_3^5 b_5   b_0^3+560 b_2^2 b_3 b_4^2 b_5 b_0^3-630 b_2 b_3^3 b_4 b_5 b_0^3\nn\\
                &&+16 b_2^4 b_4^3 b_0^2-4 b_2^3 b_3^2 b_4^2 b_0^2+108 b_2^5 b_5^2 b_0^2+16 b_2^3 b_3^3 b_5 b_0^2-72 b_2^4 b_3 b_4 b_5 b_0^2\nn
\ea
which subsequently should be written in terms of $a_i$'s. This can be factorised~\cite{Callaghan:2011jj}
to give the homologies of the singlet fields $\theta_{ij}$.

\subsection{Flipped $SU(5)$}

Flipped $SU(5)$ can  naturally emerge  in the context of
F-theory~\cite{King:2010mq,Kuflik:2010dg,Jiang:2009za}. This can be easily noticed
in the spectral cover approach where the second $SU(5)$ in the chain ${\cal E}_8\ra SU(5)\times U(5)_{\perp}$
breaks to $U(1)_X\times SU(4)$
\[
{\cal E}_8\ra SU(5)\times U(5)_{\perp}\ra\left[SU(5)\times U(1)_X\right]\times SU(4)
                        \ra\left[SU(5)\times U(1)_X\right]\times U(1)^3
\]
 $U(1)_X$ can be chosen to accommodate part of the
hypercharge while  monodromies may be  imposed among the remaining abelian factors $U(1)^3\subset SU(4)$.
The $SO(10)={\cal E}_5$ embedding of $SU(5)\times U(1)_X$ can be easily detected through the following
${\cal E}_8$ breaking pattern
\[
{\cal E}_8\supset {\cal E}_5\times SU(4)
    \ra [SU(5)\times U(1)_X]\times SU(4)\ra [SU(5)\times U(1)_X]\times U(1)^3
\]
The adjoint representation of ${\cal E}_8$ then has the $SO(10)\times SU(4)$ and successively the
$SU(5)\times SU(4)\times U(1)_X$ decomposition given by
\ba
248&\rightarrow&(45,1)+(16,4)+(\ov{16},\bar{4})+(10,6)+(1,15)\nonumber \\
&\rightarrow& (24,1)_0+(1,15)_{0}+(1,1)_{0}+(1,4)_{-5}+(1,\bar{4})_{5}+(10,4)_{-1}+(10,1)_{4}\nonumber \\
 &&+(10,\bar{4})_{1}+(10,1)_{-4} +(\bar{5},4)_{3}+(\bar{5},6)_{-2} +(5,\bar{4})_{-3}+(5,6)_{2}
\ea
In flipped $SU(5)$ we have the following accommodation of fields. The chiral matter fields,
-as in the ordinary $SU(5)$-  constitute the three components of the $16\in SO(10)$, ($16=10_{-1}+\bar 5_3+1_{-5}$
 under the $SU(5)\times U(1)_X$ decomposition). However,  the definition of the hypercharge includes a component of the
external $U(1)_X$ in such a way that flips the positions of $u^c,d^c$ and $e^c,\nu^c$,
while leaves the remaining unaltered. Indeed, employing the hypercharge definition $Y=\frac 15\left(x+\frac 16 y\right)$
where $x$ is the charge under the $U(1)_X$ and $y$ the diagonal generator in $SU(5)$,  we obtain the following
`flipped' embedding of the SM representations
\ba
F_i&=&10_{-1}\;=\;(Q_i,d^c_i,\nu^c_i)\nn\\
\bar f_i&=&\bar 5_{+3}\;=\;(u^c_i,\ell_i)\\
\ell^c_i&=&1_{-5}\;=\;e^c_i\nn
\ea
The Higgs fields are found in
\ba
H\equiv 10_{-1}\;=\;(Q_H,D_H^c,\nu_H^c)&,&\bar H\equiv \ov{10}_{+1}\;=\;(\bar Q_H,\bar D_H^c,\bar \nu_H^c)\\
h\equiv 5_{+2}\;=\;(D_h,h_d)&,&\bar h\equiv \bar 5_{-2}\;=\;(\bar D_h,h_u)
\ea
There is a remarkable fact in the flipped $SU(5)$ model, which is going to be crucial for the viability
in the F-theory construction: we observe that matter antifiveplets  carry different $U(1)_X$ charges
 from the Higgs anti-fiveplets, thus they are distinguished from each  other. Consequently,  they
do not contain exactly the same components. Several $R$-parity violating
terms are not allowed because of this distinction.

For rank one mass textures these couplings predict $m_t=m_{\nu_{\tau}}$ at the GUT scale.
However, in contrast to the standard $SU(5)$ model, down quarks and lepton mass matrices are not
related, since at the $SU(5)\times U(1)_X$ level they originate from different Yukawa
couplings.  Indeed,  the mass terms descend from the following
 $SU(5)\times SU(4)\times U(1)_X$ invariant trilinear couplings
\ba
{\cal W}_d&=&10_{-1}\cdot 10_{-1}\cdot 5^h_{2}\;\ra\; Q_i\,u_j\,h_d\\
{\cal W}_u&=&10_{-1}\cdot\bar 5_{3}\cdot\bar 5^{\bar h}_{-2}\;\ra\;Q\,u^c\,h_u+\ell\nu^c\,h_u\\
{\cal W}_l&=&1_{-5}\cdot\bar 5_{3}\cdot 5^h_2\;\ra\; e^c\,\ell\,h_d
\ea
This gives the opportunity to obtain a correct fermion mass hierarchy at
$M_W$\footnote{E.g., we can evade the naive $M_{GUT}$-mass matrix relation $m^0_{down}=m^0_{lepton}$ of the minimal
$SU(5)$  GUT. We know that in order to obtain the observed lepton and down quark mass spectrum at low energies,
at the GUT scale we should have the relations $m_{\tau}^0\approx m_b^0$, $m_{\mu}^0\approx 3\,m_s^0$
and $m_{e}^0\approx 1/3\,m_d^0$.}. Moreover,
a higher order term providing Majorana masses  for the right-handed neutrinos can be written
\ba
{\cal W}_{\nu^c}&=&\frac{1}{M_S}\ov{10}_{\bar H}\ov{10}_{\bar H}\,10_{-1}\,10_{-1}
\label{Maj}
\ea
In the present context,  the above terms descend from the following
 $SU(5)\times SU(4)\times U(1)$ invariant trilinear couplings
\ba
{\cal W}_{down}&\in&(10,4)_{-1}\cdot (10,4)_{-1}\cdot (5, 6)_{2}\\
{\cal W}_{up}&\in&(10,4)_{-1}\cdot (\bar 5,4)_{3}\cdot (\bar 5, \bar 6)_{-2}\\
{\cal W}_{\ell}&\in& (1,4)_{-5}\cdot (\bar 5,4)_3\cdot (5,6)_2
\ea
Further, in general the following Higgs terms can be written
\ba
H_iH_ih_j+\bar H_i\bar H_i\bar h_j
\ea
When $H_,\,\bar H_i$ acquire vevs, one obtains mass terms for the colour triplets
\ba
\langle H_i\rangle d^c_{H_i}D_j+\langle \bar H_i\rangle \bar d^c_{H_i}\bar D_j
\ea
As we have explained in previous sections, the abelian symmetries descending from
the breaking of $SU(4)\ra U(1)^3$ prevent  tree-level couplings for
the third generation, thus as in the case of $SU(5)$ we need to appeal to  monodromies among the $U(1)$'s.
Given that for the flipped model the highest  accompanying symmetry is $SU(4)$, there are three
possible choices for the monodromy group, namely $S_3$, ${\cal Z}_2\times {\cal Z}_2$ and
${\cal Z}_2$. The first two cases reduce the number of the available matter curves to two.
The ${\cal Z}_2$ case gives exactly three matter curves.
In the first two cases at least two families should reside on the same matter curve. Hierarchy
is then generated by flux effects~\cite{Cecotti:2009zf,Aparicio:2011jx,Camara:2011nj}.
If we wish to accommodate the families on different matter curves,
 only the ${\cal Z}_2$ monodromy allows the possibility of distinct localization of the three families.

As an example, let us see how matter curves are organised in  the case of ${\cal Z}_2$ monodromy.
Assuming  $t_i,i=1,\dots,4$, with $\sum_{i=1}^4t_i=0$ we have the following correspondence
between $t_i$ and the representations\footnote{ Note that, although $6\equiv \bar 6$, we `distinguish'
them under the weights $t_i$ $6/\bar 6\ra\pm (t_i+t_j)$. This will only result to a relabeling of the curves since
 $t_i+t_j=-(t_k+t_l)$ where all $i,j,k,l$ differ.}
\[4\ra t_i,\;
\bar 4\ra -t_i,\;6\ra t_i+t_j,\;i\ne j,\;
\bar 6\ra -(t_i+t_j),\;i\ne j
\]
Tables~\ref{M5x1} and~\ref{H5x1} show  the flipped content for the case of ${\cal Z}_2$ monodromy.
The resulting fermion mass textures and other phenomenological issues are discussed in~\cite{King:2010mq}.
\begin{table}[!t]
\centering
\renewcommand{\arraystretch}{1.2}
\begin{tabular}{|c|c|c|}
\hline
$F\in 10^j,j=1,2,3$&$\bar f\in \bar 5^j,j=1,2,3$&$\ell^c\in 1^j,j=1,2,3$\\
\hline
$(10,4)_{-1}:
\left\{\begin{array}{ll}
10^{(1)}_{-1}:&\{t_1,t_2\}\\
10^{(2)}_{-1}:&\{t_3\}\\
10^{(3)}_{-1}:&\{t_4\}\end{array}\right. $&
$(\bar 5,4)_{3}:\left\{\begin{array}{ll}
\bar 5^{(5)}_{3}:&\{t_1,t_2\}\\
\bar 5^{(6)}_{3}:&\{t_3\}\\
\bar 5^{(7)}_{3}:&\{t_4\}\end{array}\right.$&
 $(1, 4)_{-5}:\left\{\begin{array}{ll} 1_{-5}^{c(1)}:&\{t_1,t_2\}\\
                 1_{-5}^{c(2)}:&\{t_3\}\\
                 1_{-5}^{c(3)}:&\{t_4\}\end{array}\right. $\\
\hline
\end{tabular}
 \caption{Matter curves (labeled by the $SU(4)$ weights $t_i$, where $\sum_it_i=0$),
  available to accommodate the fermion generations in the case of ${\cal Z}_2$ monodromy in flipped $SU(5)$.\label{M5x1} }
\end{table}

\begin{table}[!t]
\centering
\renewcommand{\arraystretch}{1.2}
\begin{tabular}{|c|c|c|}
\hline
$\bar h\in\bar 5^i_{-2},i=1,2,3,4$&$h\in 5^{i}_{2},i=1,2,3,4$&$\theta_{ij}\in1_0^{ij}$\\
\hline
$(\bar 5 ,\bar 6)_{-2}:
\left\{\begin{array}{ll}\bar 5^{(1)}_{-2}:&\{-t_1-t_2\}\\
                     \bar 5^{(2)}_{-2}:&\{-t_3-t_4\}\\
                    \bar 5^{(3)}_{-2}:&\{-t_{1,2}-t_3\}
                    \\
                   \bar 5^{(4)}_{-2}:&\{-t_{1,2}-t_4\}\end{array}\right. $&
$( 5 , 6)_{2}:
\left\{\begin{array}{ll} 5^{(1)}_{2}:&\{t_1+t_2\}\\
                      5^{(2)}_{2}:&\{t_3+t_4\}\\
                     5^{(3)}_{2}:&\{t_{1,2}+t_3\}
                    \\
                   5^{(4)}_{2}:&\{t_{1,2}+t_4\}\end{array}\right.$&
 $1^{ij}:\{t_i-t_j\} $\\
\hline
\end{tabular}
 \caption{Higgs curves, and their labeling under the four $SU(4)$ weights $t_i$ \label{H5x1}. }
\end{table}

\section{Gauge coupling unification in F-theory models}

The spectrum of the minimal supersymmetric extension of the Standard Model is consistent with a
 gauge coupling unification at a scale $M_{GUT}\sim 2\times 10^{16}$ GeV.
In the simplest case, the SM gauge symmetry is embedded in the $SU(5)$  GUT
with  the SM  matter content incorporated  into $SU(5)$ multiplets.
However, in a string derived $SU(5)$ model,
one must confront  the mismatch  between $M_{GUT}$ and the natural gravitational scale
 $M_{Pl}\sim 1.2 \times 10^{19}$ GeV.
 We have pointed out earlier, that
in F-theory it is possible to decouple  gauge dynamics from gravity by restricting
to compact surfaces $S$ that are of del Pezzo type. The exact determination of the
GUT scale however, may depend on the spectrum and other details of the chosen
gauge symmetry and on the particular model.   In F-theory
  $SU(5)$ we are examining here, there are several sources of threshold effects  that have to be taken into account~\cite{Donagi:2008kj,Blumenhagen:2008aw,Palti:2009,Leontaris:2009wi,Leontaris:2011pu,Heckman:2011hu,Dolan:2011aq}.
Thus, we encounter thresholds related  to the flux mechanism  which  induce splitting of the gauge
couplings at the GUT scale~\cite{Donagi:2008kj,Blumenhagen:2008aw}.
A second source concerns threshold corrections generated from heavy
KK massive modes~\cite{Donagi:2008kj,Leontaris:2011pu}.  Furthermore,  corrections
to gauge coupling running  arise due to the
appearance of  probe D3-branes  generically present in F-theory compactifations
and filling  the $3+1$ non-compact dimensions while sitting  at certain points
of the internal manifold~\cite{Heckman:2011hu}.

We focus here on two sources of thresholds, namely the ones induced by fluxes and those by KK-modes.
Thresholds induced by the flux mechanism have been extensively analysed
in recent literature~\cite{Donagi:2008kj,Blumenhagen:2008aw,Leontaris:2009wi}.
 There, it   was shown that the $U(1)_Y$-flux induced splitting is compatible with the
 GUT embedding of the minimal supersymmetric standard model, provided that no
 extra matter other than  color triplets is present in the spectrum.
Thresholds originating from KK-massive modes have been discussed in~\cite{Donagi:2008kj} and
were found to be related to a topologically invariant quantity, the Ray-Singer
analytic torsion~\cite{Ray:1973sb}.  In F-theory,  KK-massive modes exist for both the gauge and
the matter fields. Taking also into account that several low energy effective models
 involve exotics in the light spectrum, it is possible that they might threaten the gauge
 coupling unification.
Here it will be  argued that under  reasonable assumptions for the matter curve
bundle structure, in a class of $SU(5)$ models the KK-massive modes do not have any effect
on the unification\cite{Leontaris:2011tw}. Alternatively, one may  implement
the requirement of unification to constrain  thresholds from KK
modes of $SU(5)$ gauge and matter field~\cite{Donagi:2008kj,Leontaris:2009wi,Leontaris:2011pu,Dolan:2011aq}.

We start with the $SU(5)$ gauge multiplet under (\ref{24Ri}) and recall the fact that
the massless exotics  $R_{\pm 5/6}$ have been eliminated by imposing the condition $ \chi(S,L^{5/6})=0$
(see eq.~\ref{noexotic}).
At the one-loop level we write
\be
\label{gauge}
\frac{16\pi^2}{g^2_a(\mu)}=\frac{16\pi^2
k_a}{g^2_s}+b_a\log\frac{\Lambda^2}{\mu^2}+\mathcal{S}_a^{(g)},\quad
a=3,2,Y\,\cdot
\ee
Here  $\Lambda$ is the gauge theory cutoff scale, $k_a=(1,1,5/3)$ are the
normalization coefficients for the usual embedding of the Standard
Model into $SU(5)$,  $g_s$ is the value of the gauge coupling at the high  scale,
and $b_a$ the one-loop $\beta$-function coefficients.
The massive modes in representations (\ref{24Ri})  induce threshold effects
to the running of the gauge couplings denoted by $\mathcal{S}_a^{(g)}$.
These can be written~\cite{Donagi:2008kj,Leontaris:2011tw} in terms of
the Ray-Singer torsion ${\cal T}_{i}$
\begin{equation}
\label{S56_2}
{\cal S}^{(g)}_a=
\frac 43 b_a^{(g)} \left({\cal T}_{5/6}-{\cal T}_0\right)+20\,k_a{\cal T}_{5/6} \cdot
\end{equation}
We absorb the term  proportional to $k_a$ into  a redefinition of  ${g}_s$ while the remaining
part suggests that we  can define $M_{GUT}$ as~\cite{Leontaris:2011pu}
\begin{equation}
\label{MGUT1}
M_{GUT}=e^{2/3\left({\cal T}_{5/6}-{\cal T}_0 \right)}\,M_C\,\cdot
\end{equation}
Here we have associated the world volume factor $V_S^{-1/4}$ with the characteristic
F-theory compactification scale $M_C$.

Next we will consider contributions arising from  chiral matter and the Higgs
fields  transforming under the standard  $10,\overline{10}$ and $5,\bar 5$ non-trivial representations.
 We should mention that the $U(1)_Y$-flux introduced in order to break   $SU(5)$ might eventually lead to incomplete $SU(5)$ representations, spoiling thus the gauge coupling unification.  However, in the previous sections we have already discussed realistic cases where the matter fields add up to complete $SU(5)$ multiplets, so that the $b_a^x$-functions contribute in proportion to the coefficients $k_a$.
 Under the above assumptions, we may write threshold terms for the KK-states leaving in
(\ref{105split}) representations  as follows~\cite{Leontaris:2011pu}
\ba
S_a^{\bar 5}&=&-\frac{4}{3}\,\beta_a\,({\cal T}_{-1/2}-{\cal T}_{1/3})+k_a\,( 2\cdot{\cal T}_{-1/2})
\label{TC5}
\\
S_a^{10}&=&+\frac{4}{3}\,\beta_a\,({\cal T}_{-2/3}-{\cal T}_{1/6})+k_a\,(6\cdot {\cal T}_{1/6})
\label{TC10}\cdot
\ea
with $\beta_a=\beta_{3,2,1}=\{\frac 32,0,1\}$ while  ${\cal T}_{q_i}$ is the torsion and the indices refer
 to hypercharges.  We now observe  that the  hypercharge  differences not proportional to $k_a$
 in both $\Sigma_{10}$ and $\Sigma_{\bar 5}$   satisfy the same condition $q_i-q_j=-\frac 56$.
 Given this property and the fact that the torsion is a topologically invariant quantity,
 one could assume the existence of bundle structures for $\Sigma_{10}$ and $\Sigma_{\bar 5}$
so that the above differences vanish. Then, only the terms proportional to $k_a$
remain which can be absorbed in a redefinition of the gauge coupling at $M_{GUT}$.

We will assume that matter resides on at most genus one ($g=1$) matter curves
 (see example discussed in section 4.1 as well as in~\cite{Beasley:2008kw})  with chiral
 matter forming complete
  $SU(5)$ multiplets. For $g=1$ in particular,
  according to Ray-Singer\cite{Ray:1973sb,Leontaris:2011tw},  the analytic torsion is
\ba
{\cal T}_{v}\equiv {\cal T}_{z=u-\tau v}&=&\ln\left|\frac{e^{\pi\,i\tau\,v^2}\vt_1(u-\tau v,\tau)}{\eta(\tau)}\right|
\label{TR1}
\ea
For $v\to v-1$, making use of known theta-function identities we observe
\ba
{\cal T}_{v-1}\equiv {\cal T}_{z=u-\tau (v-1)}&=&\ln\left|\frac{e^{\pi\,i\tau\,(v-1)^2}\vt_1(u-\tau( v-1),\tau)}{\eta(\tau)}\right|= {\cal T}_{z=u-\tau v}\cdot \label{TR2}
\ea
In order to use this result, we need to make a proper identification of  the hypercharge $q_i$.
Considering now two successive hypercharge values $q_i,q_j$ such that $|q_i-q_j|=\frac 56$ and using the association
\ba
v_i&=&\frac{q_i}{|q_i-q_j|}
\label{Yv}
\ea
we get the identification
\[{\cal T}_{u-\tau v_i}\leftrightarrow{\cal T}_{q_i} \cdot \]
With this embedding we can easily see that the differences ${\cal T}_{-2/3}-{\cal T}_{1/6}=0 $
and ${\cal T}_{-1/2}-{\cal T}_{1/3}=0$  so that  threshold corrections vanish and unification is retained.
Thus, adding matter contributions  of complete $SU(5)$ representations to (\ref{gauge}),
while assuming that the Higgs-triplet pair decouples at $M_X$  we finally get
\begin{equation}
\label{FinMG}
\begin{split}
\frac{16\pi^2}{g^2_a(\mu)}&=\frac{16\pi^2 }{g^2_G}\,k_a+
(b_a^{(g)}+b_a)\log\frac{M_{GUT}^2}{\mu^2 }
+b_a^T\log\frac{M_{GUT}^2}{M_X^2 }
\end{split}
\end{equation}
where $b_a^T, b_a$ MSSM beta functions with and without the triplet-pair contribution and the
GUT value of the gauge $g_G$  coupling is related to the string $g_s$ coupling by
\[\frac{16\pi^2 }{g^2_G}=\frac{16\pi^2 }{g^2_s}+20{\cal T}_{5/6}+6{\cal T}_{1/6}+2{\cal T}_{1/3}\cdot\]

\section{Summary and recent progress}

 In the previous sections we have presented  techniques for the construction of F-theory $SU(5)$ models
  and analysed ways and novel mechanisms for symmetry breaking and doublet triplet splitting.  In F-theory,
   important properties of the effective field theory model depend on the specific geometry of the compact space
   and the internal fluxes.   Thus, we have investigated how
  the triplet-doublet splitting problem for example can be solved by judicious  choice of fluxes in
  order to split the $SU(5)$ Higgs fiveplets~\cite{Beasley:2008kw,Marsano:2009wr,Dudas:2010zb,Leontaris:2010zd}.

  Several other important issues of GUT models have been successfully treated in their  F-theory analogues.
  Thus, it has been suggested~\cite{Hayashi:2009bt} that unwanted proton decay operators  can be avoided through  the  incorporation of an R-symmetry  by invoking  symmetries of the manifold and of the fluxes~\cite{King:2010mq,Dudas:2010zb}. Alternative ways have also been presented based on the abelian factors~\cite{Grimm:2010ez,Ludeling:2011en,Kuflik:2010dg,Krause:2011xj} and the extension of
  the $SU(5)$ gauge group to flipped $SU(5)\times U(1)_X$ discussed in the previous section.

Further progress has also been made towards the computation of the Yukawa couplings and the determination of the fermion mass spectrum\cite{Aparicio:2011jx,Camara:2011nj,Palti:2012aa}. We have already explained that in F-theory chiral matter is localised along the intersections of the surface $S$ with other 7-branes, while  Yukawa couplings are formed when three of these curves intersect at a single point on $S$.  Their computation relies on the knowledge of the profile of the wavefunctions of the states participating in the intersection. When a specific geometry is chosen for the internal space (and in particular for the GUT surface) these profiles are found by solving the corresponding equations of motion~\cite{Beasley:2008dc}.  Besides, the precise knowledge of the common gauge coupling value at the GUT scale is crucial for  the determination of the Yukawa couplings involved in the calculation of the fermion mass
spectrum~\cite{Leontaris:2011pu}.
Then, their values are obtained  by  computing the integral of the overlapping wavefunctions  at the triple intersections.  Despite this important success towards a reliable computation of undetermined
parameters of GUTs and the Standard Model, yet a lot of work is required to formulate a complete
picture of an F-theory derived effective low energy model; because a theory, no matter how beautiful it is, has to face the relentless test of the experimental proof.

\vfill
{\bf Acknowledgement.}  Research partially
supported by the EU ITN grant UNILHC 237920 (Unification in the LHC era).

\end{document}